\documentclass[12pt,A4,preprint,superscriptaddress,aps]{revtex4}

\usepackage[english]{babel}

\usepackage{amssymb}
\usepackage{graphicx}
\usepackage{epsfig}

\def\dkkk{ D^+ \to K^- K^+ K^+}

\def\lp {\left( }
\def\rp {\right) }
\def\lb {\left[ }
\def\rb {\right] }
\def\lc {\left\{ }
\def\rc {\right\} }
\def\ra {\rangle }
\def\la {\langle }

\def\rar {\rightarrow}
\def\lrar {\leftrightarrow}

\def\beq{\begin{equation}}
\def\eeq{\end{equation}}
\def\bea{\begin{eqnarray}}
\def\eea{\end{eqnarray}}
\def\nn {\nonumber}

\def\cd {\!\cdot\!}

\def\ct {\tilde{c}}

\def\dr {\partial }

\def\db {\bar{d}}
\def\ub {\bar{u}}
\def\sb {\bar{s}}
\def\Ob {\bar{\Omega}}

\def\Kb {\bar{K}}

\def\rtw {\sqrt{2}}
\def\rth {\sqrt{3}}
\def\rts {\sqrt{6}}
\def\sp {\!+\!}
\def\sm {\!-\!}
\def\se {\!\!=\!\!}

\def\mr2 {m_\rho^2 }

\def\cK {{\cal{K}}}

\def\cL {{\cal{L}}}

\def\a{\alpha}
\def\b{\beta}
\def\d{\delta}
\def\e{\epsilon}
\def\g{\gamma}
\def\Gb {\bar{\Gamma}}
\def\D {\Delta}
\def\f {\phi}
\def\G {\Gamma}

\def\l{\lambda}
\def\m{\mu}
\def\n{\nu}
\def\o{\omega}

\def\p {\pi}

\def\r{\rho}
\def\s{\sigma}

\def\S{\Sigma}

\def\x {\chi}
\def\y {\eta}
\def\z {\zeta}
\def\w{\omega}

\begin{document}

\title{Multi-Meson Model for the $D^+\to K^+K^-K^+$ decay amplitude }




\author{R.T. Aoude}
\affiliation{Centro Brasileiro de Pesquisas F\'isicas, Rio de Janeiro, Brazil}
\affiliation{PRISMA Cluster of Excellence and Mainz Institute for Theoretical Physics, 
Johannes Gutenberg-Universit{\"a}t Mainz, Germany}
\author{P.C. Magalh\~aes}
\email[]{patricia.magalhaes@tum.de}
\affiliation{Centro Brasileiro de Pesquisas F\'isicas, Rio de Janeiro, Brazil}
\affiliation{Technical University of Munich, Germany}
\author{A.C. dos Reis}
\affiliation{Centro Brasileiro de Pesquisas F\'isicas, Rio de Janeiro, Brazil}
\author{M.R. Robilotta}
\affiliation{Instituto de F\'isica,  Universidade de S\~ao Paulo, S\~ao Paulo, Brazil}

\date{\today }
\preprint{MITP/18-041}

\begin{abstract}
We propose a novel approach to describe the $\dkkk$ decay
 amplitude, based on chiral effective Lagrangians,  which can be used  
to extract information about $K\bar{K}$ scattering.
Our trial function is an alternative to  the widely used isobar model and
includes both  nonresonant three-body interactions and
two-body rescattering amplitudes, based on coupled channels and resonances, 
for S- and P-waves with isospin $0$ and $1$.
The latter are unitarized in the $K$-matrix approximation and represent the only
source of complex phases in the problem.  
Free parameters are just resonance masses and coupling constants, 
with transparent physical meanings. 
The nonresonant component, given by chiral symmetry as a real polynomium,
is an important prediction of the model, which goes beyond the (2+1) approximation.
Our approach allows one to disentangle the two-body scalar contributions with 
different isospins, associated with the $f_0(980)$ and  $a_0(980)$ channels.
We show how the $K\bar{K}$ amplitude can be obtained from the decay $\dkkk$
and discuss extensions to other three-body final states.
\end{abstract}

\pacs{...}

\maketitle

\section{introduction}
 Nonleptonic weak decays of heavy-flavoured mesons  are extensively 
used in light meson spectroscopy. 
Owing to a rich resonant structure, these decays 
provide a natural place to study hadron-hadron interactions at low energies. 
In particular, almost 20 years ago,  three-body decays of charmed mesons could 
confirm the existence of the controversial 
scalar states $f_0(600)$ (or sigma)\cite{sigma} and  kappa$(800)$\cite{E791kappa}.
More comprehensive investigations can be done nowadays, 
using the very large and pure samples provided by the LHC experiments, 
and still more data is expected in the near future,  with Belle II experiments.

Three-body hadronic decays of heavy-flavoured mesons involve combinations of different classes 
of processes, namely  heavy-quark weak
transitions, hadron formation and final-state interactions (FSI), whereby the hadrons 
produced in the primary vertex are allowed to interact in many different ways before being detected. 
Final-state processes include both proper three-body interactions 
and a wide range elastic and inelastic coupled channels, involving resonances.
In this framework,  a question arises, concerning 
how to obtain information about  two-body scattering amplitudes from the abundant
data on three-body systems.

The key issue of this program is the modeling of the decay amplitudes.
Most amplitude analyses have been performed
using the so-called isobar model, in which the decay amplitude is represented by a
coherent sum of both nonresonant and resonant contributions.
This approach, albeit largely employed \cite{LHCb_BW_exemplo}, has
conceptual limitations. 
The outcome of isobar model analyses are resonance parameters such as fit fractions, masses and widths, 
which are neither directly related to any 
underlying dynamical theory nor provide clues to the identification of
two-body substructures.
Thus, the systematic interpretation of the isobar model results is rather difficult.

This situation motivated in the past decade efforts towards building models that are
based on more solid theoretical grounds.
Those models improve essentially the two-meson interaction
description in the FSI, with the use of dispersion relations and chiral perturbation theory. 
Most of them work in the quasi-two-body (2+1) approximation, where interactions with the third 
particle are neglected. 
Recently, a collection of parametrizations based on analytic and unitary meson-meson 
form factors for D and B three-body hadronic decays within the (2+1) approximation  
was presented in Ref.\cite{BoitoRESUM}. 
Three-body FSIs were also considered and, in particular,
shown to play a significant role in the $D^+ \rar  K^- \pi^+ \pi^+$  decay.
In this process, three-body unitarity was implemented in different ways, by means of 
Faddeev-like decompositions\cite{BR,PatWV,tobias}, Khuri-Treiman equation\cite{kubis} 
or triangle diagrams \cite{satoshi}. 
Whilst differing in methods and techniques, all these theoretical
efforts have in common the attempt to include, in a systematic way,  knowledge of two-body 
systems in the description of the decay amplitudes.

This work departs from the same broad perspective, but concentrates explicitly 
on the derivation of two-body scattering amplitudes 
from three-body decays. 
In order to do so, we suggest a new approach based on effective Lagrangians and apply it to the
 $\dkkk$ decay.
This process is interesting because there is very little information
 available on kaon-kaon scattering, regarding both theory and experiment. 
 Concerning the latter, 
one only has access to $\p\p$ elastic scattering data \cite{Hyams} 
 and to the inelastic channel $\p\p\to K\bar{K}$\cite{Hyams, cohen}.
 Information about $K\bar{K}$
interaction can be estimated by imposing unitarity constraints on the  $\p\p$ data.
On the theory side, $K\bar{K}$ amplitudes have been calculated in next-to-leading order 
chiral perturbation theory.  
Aiming at a full coupled-channel description, it was extended up to 1.2 GeV,
using form factors \cite{mousallam_KKff} 
to describe the $\eta\pi \to K\bar{K}$ contribution to
 $\eta\to \p\p\p$ decay\cite{mousallam_eta3pi}, 
 or unitarized ressummation  
techniques\cite{OllerMeissner}, 
to include $\p\pi \to K\bar{K}$ in the context of FSI of
 $J/\Psi \to \phi \p\p(K\bar{K})$ decays.

The main purpose of this work is disclose
 information about the dynamics of $K\bar{K}$ interactions by disentangling 
the two-body  contributions contained in the $\dkkk$ amplitude. 
In our model, the description of the $K\bar{K}$ interaction relies on a chiral Lagrangian with resonances, 
including all possible coupled channels for ($J=0,1;I=0,1$), extended to non-perturbative regimes 
 by means of unitarization.
A relevant feature of the model is that the relative contribution and phase of each 
 component is fixed by theory, and this represents an important difference 
 with the isobar model. Although the formalism is developed for a specific process it
can be useful in other decays into three kaons. 

This paper extends and supersedes a previous version\cite{TM} and is organized as follows.
The motivation for building the amplitude is discussed in section II, whereas the model  is presented 
in sections III and IV. 
The suggested amplitude for data fitting, together with a comparison between scattering 
and decay amplitudes is discussed in section V. 
Some simulations and general remarks are given in section VI. 
Details of the calculations are given in the appendices.
\\


\section{motivation for a new model}
\label{motivation}

The  isobar model, widely used for describing heavy-meson decays into three pseudoscalars,
relies on the assumption that these processes are dominated by 
intermediate states involving a spectator plus a resonance,  and also includes non-resonant contributions.
In the decay $ H \rar P_1 P_2 P_3 $, of a heavy meson $H$ into three pseudoscalars $P_i$,
the isobar model emphasizes the sequence $H\rar R\, P_3$, followed 
by $R \rar P_1 P_2$.

The full $H\rar P_1 P_2 P_3 $ decay amplitude is denoted by $T$ and  
the isobar model employs a guess function to be fitted to data in the form of the coherent sum
\bea 
&& {T} = c_{nr} \; \tau_{nr} +  \sum c_k \; \tau_k \;,
\label{mot.1}
\eea
the subscript $nr$ referring to the non-resonant term and the label $k$ associated with
resonances, as many of them as needed.
The coefficients $c_k=e^{i\theta_k}$ are complex parameters, to be determined by data.
The choice $\tau_{nr}=1$ is  usual for the non-resonant term, whereas the
sub amplitudes $\tau_k$ depend on the invariant masses of the problem. 
For each resonance considered, the function $\tau_k$ is given by
$  \tau_k = [FF] \times [\, \mathrm{angular} \; \mathrm{factor} \,] 
\times [\mathrm{line} \; \mathrm{shape} ]_k $,
where ${FF}$ stands for form factors, the angular factor is associated with angular momentum channels,
and $ [\mathrm{line} \; \mathrm{shape} ]_k$ represents a resonance line shape, described by either  
a Breit-Wigner function such as
$(BW)_k = 1/[s- m_k^2 + i\,m_k\, \Gamma_k]$, $m_k$ and $\Gamma_k$ being the 
resonance mass and width, or by variations, such as the Flatt\'e  or Gounaris-Sakurai forms.
The angular factor allows one to distinguish partial wave contributions and to employ the 
decomposition $ T = T^S + T^P + \cdots$.
 
A good fit to decay data based on the structure given by eq.(\ref{mot.1}),
would yield an empirical set of complex parameters $c_{nr}$ and $c_k$. 
However, a question arises regarding the meaning of these parameters.
Would they be useful to shed light into yet unknown two-body substructures of the problem?
Can they provide reliable information about scattering amplitudes?
If we denote two-body scattering amplitudes by $A$
this question may be restated as: can one extract $A$ directly from $T\, $?
As we argue in the sequence, answers to these questions do not favour the isobar model.

On general grounds, 
there is no direct connection between a heavy-meson decay amplitude $T$ and 
two-body scattering amplitudes $A$, involving the same particles.
Their relationship involves several issues, which we discuss below.
\\
{\bf a. dynamics - }  
The  dynamical contents of $T$ and $A$ are rather different,
since the former must include  weak vertices, which cannot be present in the latter.
Specific features of $W$-meson interactions are important to $T$ and irrelevant to $A$.
Therefore, although scattering amplitudes $A$  may be substructures of $T$, there
is no reason whatsoever for assuming that these $A$'s  are either identical or 
proportional to $T$.  
This is supported by case studies.
For instance, some time ago, the FOCUS collaboration\cite{FOCUS} produced 
a partial-wave analysis of the   $S$-wave $K^- \pi^+$ amplitude from the decay 
$D^+ \rar K^- \pi^+ \pi^+$.
Several groups then compared \cite{compare} the phase of this empirical amplitude 
directly with that from the LASS   $K^- \pi^+$  scattering data\cite{LASS}  and 
the discrepancy found was seen as a puzzle.
The fact that the FOCUS phase was negative at low energies was considered to be especially odd.
In the language of this discussion, this kind of puzzle arose just because one was trying to compare 
$T$ and $A$ directly.
The difference between observed $S$-wave decay and scattering phases was later explained
by considering meson loops in the weak sector of the problem\cite{BR, PatWV}.
These loops account for the extra phases observed.
\\
{\bf b. good quantum numbers: - } 
Isospin is broken by weak interactions and
is a good quantum number for $A$, but not for $T$.
Scattering amplitudes $A$ depend both on the angular momentum $J$ and on the
isospin $I$ of the channel considered,  whereas just a $J$ dependence 
can be extracted from an empirical decay amplitude $T$. 
This point will be recast on more technical grounds while we discuss our model.
For the time being, it suffices to stress that it 
is impossible to derive directly $A^{(J,I)}$  from $T^{(J)}$
simply because the former contains more structure than the latter.
An extraction of $A^{(J,I)}$ from $T^{(J)}$ would amount to generating physical content about the
isospin structure.
\\
{\bf c. coupled channels - }  It is well known that scattering amplitudes 
include important inelasticities  due to couplings of intermediate states.
For instance, as Hyams et al.\cite{Hyams} point out, 
$K\Kb$  intermediate states do influence elastic $\p\p$ scattering
at some energies.
Since scattering amplitudes $A$ are substructures of the decay amplitude $T$,
coupled channels present in the former must also show up in the latter. 
In general, guess functions better suited for accommodating data should have structures similar
to those used in meson-meson scattering Refs.\cite{Hyams, mousallam_KKff, Pelaez}.
In the case of the isobar model, the simple guess functions usually employed 
fail to incorporate these intermediate couplings.
\\
{\bf d. unitarity - }  Good fits to Dalitz plots data may require several 
resonances with the same quantum numbers.
At present, conceptual techniques are available which preserve unitarity while 
incorporating several resonances into amplitudes\cite{OOunit}.
This allows one to go beyond the isobar model, where the amplitude is constructed 
as sums of individual line shapes (Breit-Wigner), as in eq.(\ref{mot.1}), 
a procedure known to violate unitarity, even in the case of scattering 
amplitudes\cite{unit}.
\\
{\bf e. non-resonant term - } The non-resonant term may be important and involve
less known interactions.
In the case of  heavy meson decays and some leptonic reactions, 
available energies can be large enough for allowing the simultaneous 
production of several pseudoscalars at a single vertex.
Multi-meson dynamics then becomes relevant.
For instance, the process $e^-\,e^+ \rar 4\,\p$ involves the 
matrix element  $\la \p\p\p\p|J_\g^\m|0\ra$, $J_\g^\m$ being the
electromagnetic current\cite{EU}.
A similar matrix element, with $J_\g^\m$ replaced with the
weak current $(V\sm A)^\m$, describes the decay $\tau \rar \n \, 4\p$\cite{EU}.
Interactions of this kind are also present in the model for $ \dkkk $ we discuss here.
\\
{\bf f. lagrangians - } 
Although the point of departure of the isobar model may be sound, the problems mentioned 
tend to corrode the physical meaning of parameters it yields from fits.
Thus, even if these fits are precise, the relevance of the parameters extracted remains restricted to 
specific processes.
Moreover, in particular, one cannot rely on them for obtaining scattering information.
The most conservative way of ensuring that the physical meaning of parameters is preserved 
from process to process is to employ lagrangians, 
which rely on just masses and coupling constants.
Guess functions for heavy-meson decays constructed from lagrangians 
yield  free parameters which allow the straightforward derivation
of scattering amplitudes.

\section{dynamics}
\label{dynamics}

The fundamental QCD lagrangian for strong interactions is written 
in terms of gluons and quarks, the basic degrees of freedom.
As the theory allows for gluon self-interactions,  perturbative calculations hold at high energies only.
At present, intermediate-energy reactions cannot be described in terms of quarks and gluons,
and one is forced to rely on effective theories.
At low energies, chiral perturbation theory (ChPT) \cite{WChPT, GL84, GL85} is highly successful.
It is ideally suited for describing interactions of pseudoscalar mesons in the $SU(3)$  flavour sector, 
but can also encompass baryons.
A prominent feature of ChPT is that it realizes the hidden symmetry of the QCD ground state, 
that manifests itself as a vacuum filled with $ u\ub, d\db $, and $ s\sb $  states.
The lowest energy excitations of this vacuum are the pseudoscalar mesons, 
which are highly collective states. 
Another remarkable feature of the theory is that it yields multi-meson contact interactions.
 For instance, depending on the energy, reactions such as $ \pi \pi \rar \pi \pi K \Kb $
may involve a single interaction.
On a more technical side, in ChPT,  amplitudes are systematically 
expanded in terms of polynomials, involving both kinematical variables and quark masses.
The orders of these polynomials, assessed at a scale $\Lambda \sim 1$ GeV,
determine a dynamical hierarchy and
leading order (LO) contributions correspond to multi-meson contact interactions, 
whereas resonance exchanges are next-to-leading order (NLO).
This understanding motivated an extension  of the original chiral perturbation theory formalism, 
known as (ChPTR), in which resonances are explicitly included \cite{EGPR}. 
At present, ChPT yields the most reliable representation of 
the Standard Model at low energies.

Low-energy applications of ChPT are normally restricted to regions below the $\rho$ mass
whereas, in $D$ decays, energies above $1.5\,$GeV are involved.
Therefore, the description of hadronic interactions at those
higher energies requires further extensions of the theory,
which must include  non-perturbative effects in a controlled way. 
A widely used and rather successful approach consists in ressumming a Dyson series based on chiral interactions, 
so as to obtain unitary scattering amplitudes\cite{OOunit}.
In this work, we deal with the process $ \dkkk $ and, in principle, it should be described by
a properly unitarized three-body amplitude.
However, this is beyond present possibilities and, following the usual practice, 
we  work in the so called $(2\!+\!1)$ approximation,
in which two-body unitarized amplitudes are coupled to spectator particles.
Throughout the paper, we use the notation and conventions of Ref.\cite{EGPR}.
If needed, another extension scheme for ChPT, based on  the explicit inclusion 
of heavy mesons\cite{HM}, is also available.

\begin{figure}[h] 
\includegraphics[width=1\columnwidth,angle=0]{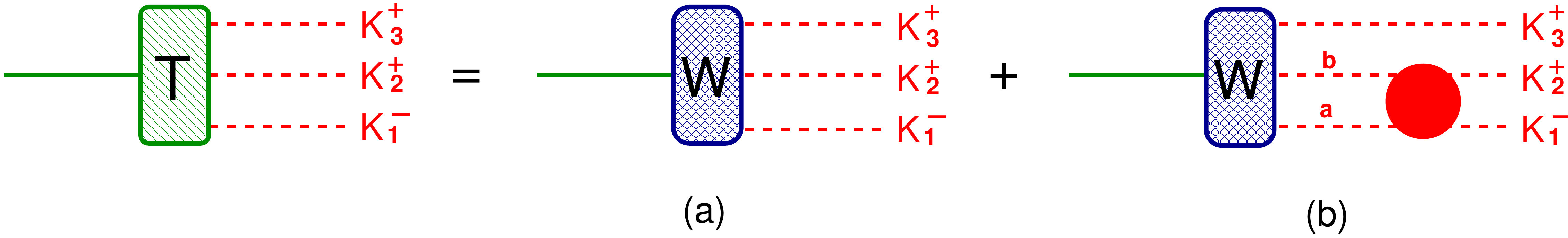}
\caption{Amplitude $T$ for $D^+ \rar K^-\,K^+\, K^+$:
(a) primary weak vertex; (b) weak vertex dressed by final state interactions;
the full line is the $D$, dashed lines are pseudoscalars.}
\label{decay}
\end{figure}

The theoretical description of a heavy meson decay into pseudoscalars 
involve two quite distinct sets of interactions.
The first one concerns the primary weak vertex, in which a heavy quark,
either $c$ or $b$, emits a $W$ and becomes a $SU(3)$ quark.
As this process occurs inside the heavy meson, 
it corresponds to the effective decay of a $D$ or a $B$ into 
a first set of $SU(3)$ mesons.
ChPT is fully suited for describing these effective processes.
The primary weak decay is then followed by purely hadronic final state interactions (FSIs), 
in which the mesons produced initially  rescatter in many different ways, 
before being detected.
The decay $\dkkk$ is doubly-Cabibbo-suppressed and any model describing it 
should involve  a combination of these  
two parts, as suggested by Fig.\ref{decay}.

In this work we allow for the  coupling of intermediate states and, within the $(2+1)$ approximation,
final state interactions are always associated with loops describing two-meson propagators.
This provides a topological criterion for distinguishing the primary weak vertex from FSIs,
namely that the former is represented by tree diagrams and the latter by a series with any 
number of loops.
Each of these loops is multiplied by a tree-level scattering amplitude $\cK$ and,
schematically, this allows the decay amplitude $T$ to be written as 
\bea 
T = (\mathrm{weak \; tree}) \,\times\,
\lb 1 + (\mathrm{loop} \times \cK) + (\mathrm{loop}\times \cK)^2 
+ (\mathrm{loop}\times \cK)^3 + \cdots \rb \,.
\label{dyn.1}
\eea
The term within square brackets involves strong interactions only and represents a geometric series
for the FSIs, which can be summed.
Denoting this sum by $S$, one has $S=1/[1-(\mathrm{loop}\times \cK)] $, which
corresponds to the model prediction for the resonance line shape.

\begin{figure}[h] 
\includegraphics[width=1\columnwidth,angle=0]{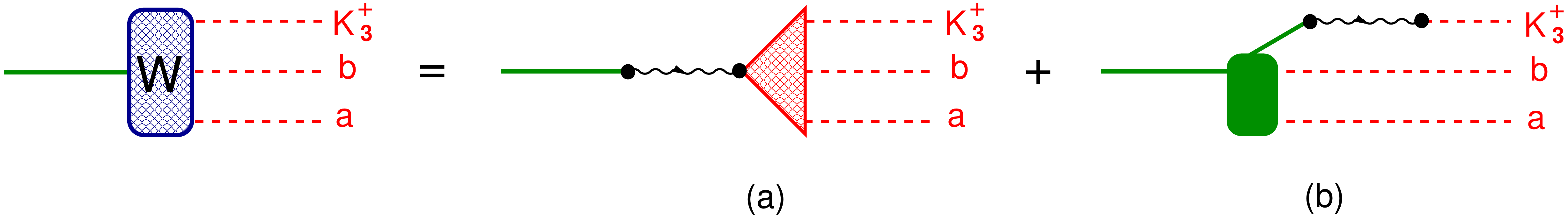}
\caption{Competing topologies for the decay $D^+ \rar K^-\,K^+\, K^+$;as
 the pair $P^a P^b$ is produced either after (a) or before (b) the weak interaction.}
\label{weak}
\end{figure}

\begin{figure}[h] 
\includegraphics[width=.7\columnwidth,angle=0]{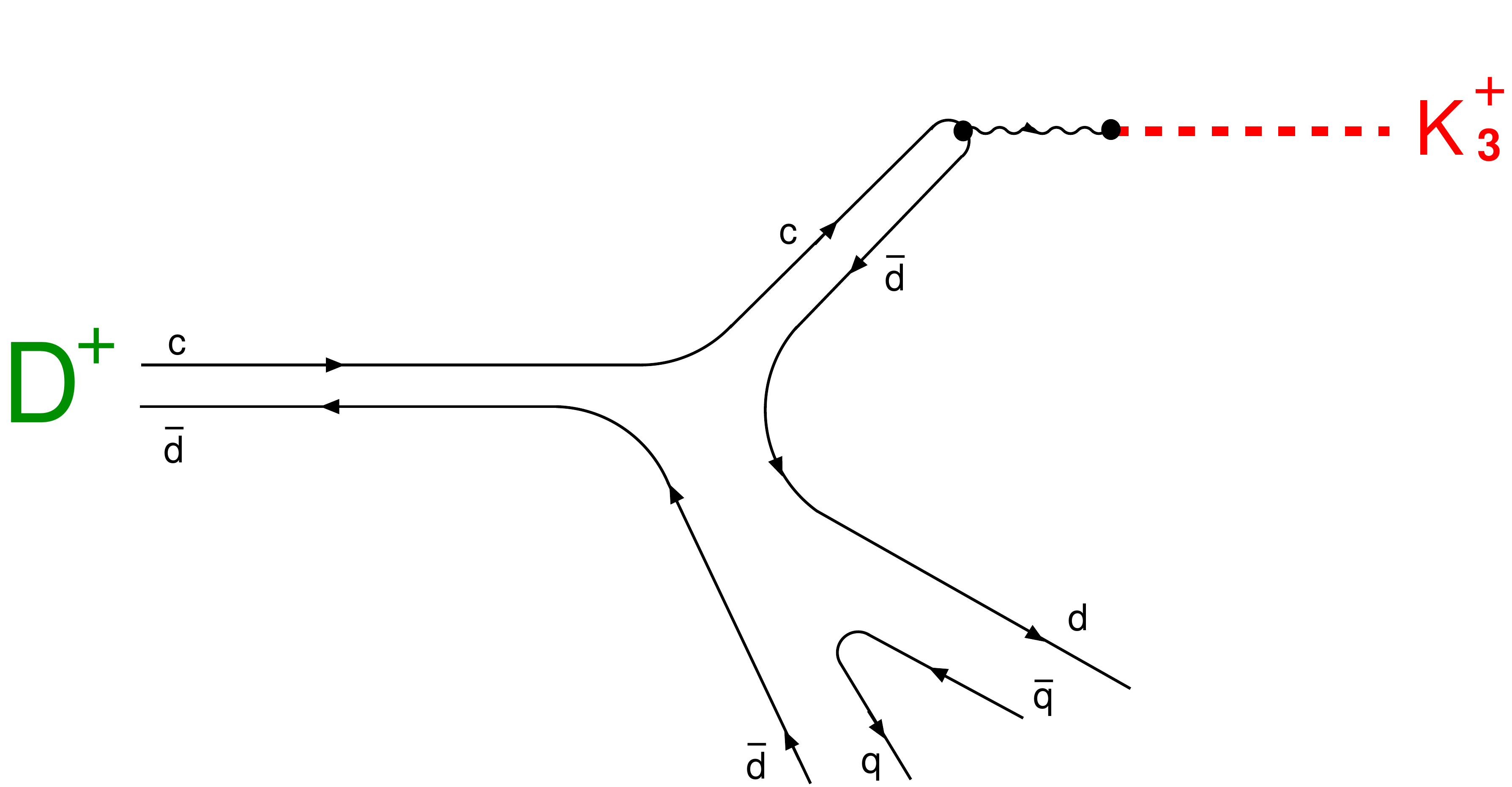}
\caption{Quark content of topology (b) of Fig.\ref{weak}.}
\label{quark}
\end{figure}

The weak amplitude describes the process $ D \rar (P^a P^b) K^+$ at tree level,
where $P^i$ corresponds to a  pseudoscalar with $ SU(3)$ label $i$. 
There are two competing topologies representing it, given by Fig.\ref{weak}.
A peculiar feature of these vertices  is that process (a) can yield $P^a P^b = K^- K^+$, 
whereas process (b) cannot.
This can be seen by inspecting the quark structure of the latter, given in Fig.\ref{quark}, 
which shows that just a $d \db$  pair is available as a source of the two outgoing mesons
at the strong vertex.
Hence one could have $P^a P^b = \p^0 \p^0, \p^+ \p^-, K^0 \Kb^0$,
but not  $P^a P^b = K^- K^+$.
The production of a $K^- K^+$ final state by mechanism (b) would thus require at least one FSI. 
In terms  of the scheme depicted in eq.(\ref{dyn.1}), this means that the first factor
within the square bracket would be absent and the decay amplitude could be rewritten as
\bea 
T = (\mathrm{weak \; tree}) \,\times\,(\mathrm{loop} \times \cK)\times
\lb 1 + (\mathrm{loop} \times \cK) + (\mathrm{loop}\times \cK)^2 
+ (\mathrm{loop}\times \cK)^3 + \cdots \rb \,.
\label{dyn.2}
\eea
Mechanism (b) is therefore suppressed when compared with mechanism (a).
The Multi-Meson-Model (Triple-M) for the $\dkkk$ amplitude proposed here
assumes the dominance of  process (a) of Fig.\ref{weak},
whereby the decay proceeds through the steps 
$D^+\to W^+ \to K^+ K^- K^+$.

\section{multi-meson-model for $D^+\to K^- K^+ K^+$}

\begin{figure}[h] 
\begin{center}
\includegraphics[width=1\columnwidth,angle=0]{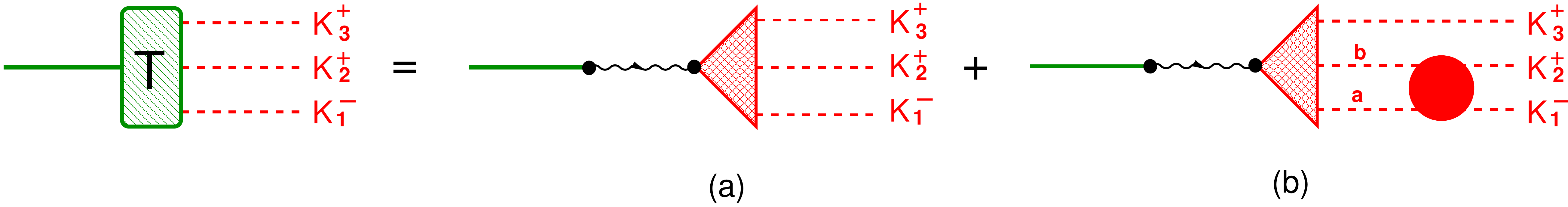}
\caption{Decay amplitude for $D^+ \rar K^-\,K^+\, K^+$; 
the weak vertex proceeds thought the intermediate steps 
$D^+ \rar W^+$  and $W^+ \rar K^-\,K^+\, K^+$ and 
strong final state interactions are encompassed by the 
scattering amplitude $A$ (full red blob).}
\label{ampT}
\end{center}
\end{figure}

Our model is based on the assumption that the weak sector of the doubly-Cabibbo-suppressed decay 
$ D^+ \rar K^- K^+ K^+ $ is dominated by the process 
shown in Fig.\ref{weak} (a), in which   quarks $ c $ and  $ \db $ in the  $ D^+ $ annihilate 
into a $ W^+ $, which subsequently hadronizes.
The primary weak decay is followed by final state interactions, involving 
the scattering amplitude $A$.
This yields the decay amplitude $T$ given in Fig.\ref{ampT}, 
which includes the weak vertex and indicates that the relationship with $A$ is not straightforward. 

%

This {\em decay} amplitude is given by 
\bea
 T =  -  \lb \frac{G_F}{\rtw} \,\sin^2\theta_C\rb \,\la K^-(p_1)\, K^+(p_2) \, K^+(p_3) |A^\m|\, 0 \,\ra  \;
\la \,0\,| A_\m |\, D^+(P) \ra \;,
\label{mmm.1}
\eea
where $ G_F $ is the Fermi decay constant, $ \theta_C $ is the Cabibbo angle, the $ A^\m $ are
axial currents and $P=p_1+p_2+p_3\,$. 
Throughout the paper, the label 1 refers to the $K^-$, the label 3 the spectator $K^+$
and kinematical relations are given in appendix \ref{kin}.

Denoting the  $ D^+ $ decay constant by $ F_D $, we write 
$ \la \,0\,| A_\m |\, D^+(P) \ra  = -i\,\rtw\,F_D \,P_\m $
and find a {\em decay} amplitude proportional to the divergence of the remaining axial current, given by
\bea
 T &\!=\!&  i\, \lb \frac{G_F}{\rtw} \,\sin^2\theta_C\rb \,\rtw \,F_D \; 
 \lb  P_\m  \, \la A^\m \ra  \rb \;,
\label{mmm.2}
\eea
with $ \la A^\m \ra=\la K^-(p_1)\, K^+(p_2) \, K^+(p_3) |A^\m|\, 0 \,\ra  $.
This result is important because, if $SU(3)$ were an exact symmetry, the axial current would 
be conserved and the amplitude $T$ would vanish.
As the symmetry is broken by the meson masses, one has the partial conservation of the axial current (PCAC)
and $T$ must be proportional to $M_K^2$.
In the expressions below, this becomes a signature of the correct implementation of the symmetry.

\begin{figure}[h] 
\includegraphics[width=1\columnwidth,angle=0]{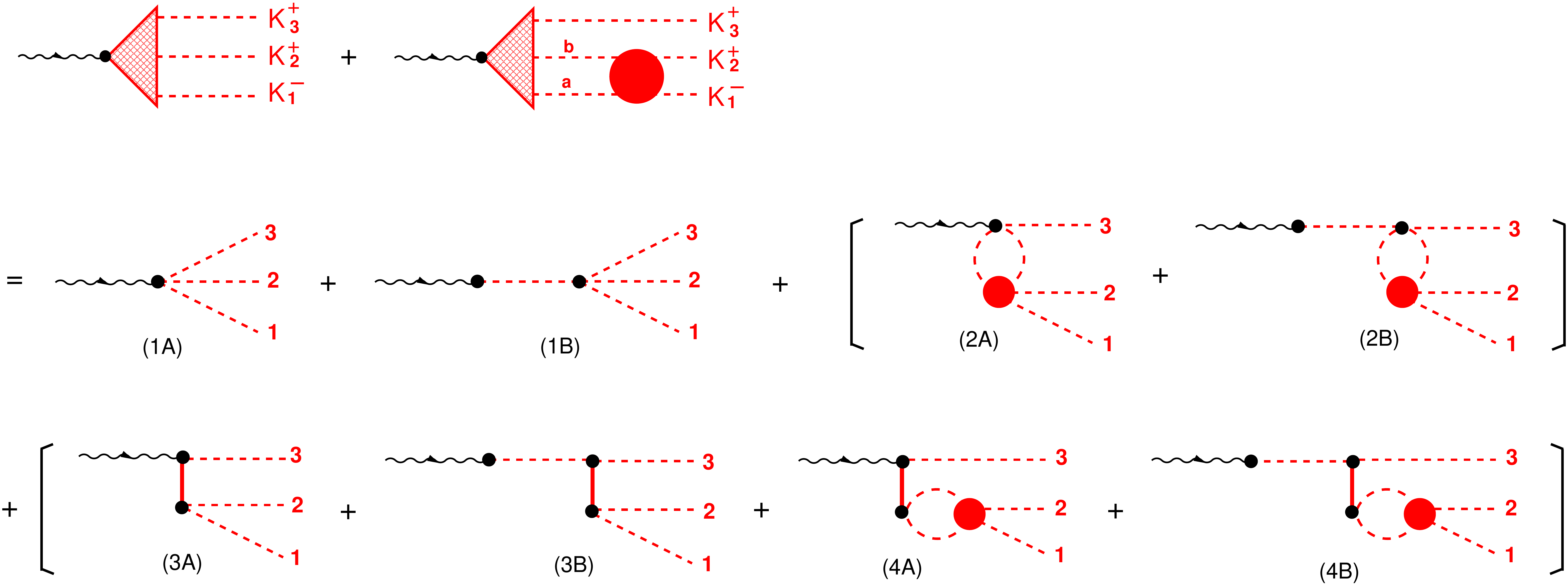}
\caption{Dynamical structure of triangle vertices in Fig.\ref{ampT};
the wavy line is the $ W^+ $, dashed lines are mesons, continuous lines are resonances
and the full red blob represent meson-meson scattering amplitudes, described in Fig.\ref{modA}; 
all diagrams within square brackets should be symmetrized, by making $ 2 \lrar 3 $. }
\label{modT}
\end{figure}

The rich dynamics of the decay amplitude $T$ is incorporated in  
the current $ \la A^\m \ra $ and displayed  in Fig.\ref{modT}.
Diagrams  are evaluated using the techniques described in Refs.\cite{GL85,EGPR}.
In chiral perturbation theory, the primary couplings of the $ W^+ $ to the $ K^-K^+K^+ $ system
always involve a direct interaction, accompanied by a kaon-pole term, denoted by (A) and (B) in the figure.
Only their joint contribution is compatible with PCAC.
Diagrams (1A+1B) are LO and describe a non-resonant term, a proper three body interaction,
which goes beyond the $ (2+1) $ approximation, whereas
Figs. (2A+2B) allow for the possibility that two of the mesons  rescatter, after being 
produced in the primary weak vertex.
Diagrams (3A+3B) are NLO and describe the {\em production} of  bare resonances 
at the weak vertex, 
whereas final state rescattering processes (4A+4B) endow them with  widths.

\subsection{two-body unitarization and resonance line shapes}
\label{denominators}

In the description of the two-body subsystem,
we consider just $S$- and $P$- waves, 
corresponding to $ (J=1,0, I=1,0) $ spin-isospin  channels.
The associated  resonances are $ \rho(770) $, $ \phi(1020) $, $ a_0(980) $, 
and two $SU(3)$ scalar-isoscalar states, $ S_1 $ and $ S_o $, corresponding
to a singlet  and to a member of an octet, respectively.
The  physical $ f_0(980) $, together with a higher mass $ f_0 $ state,  would 
be linear combinations of $ S_1 $ and $ S_o $.
Depending on the channel, the intermediate two-meson propagators may involve
$ \p\p $, $ KK $, $ \eta\eta $, and $ \p \eta $ intermediate states, 
so there is a large number of  coupled channels to be considered.

\begin{figure}[h] 
\includegraphics[width=1\columnwidth,angle=0]{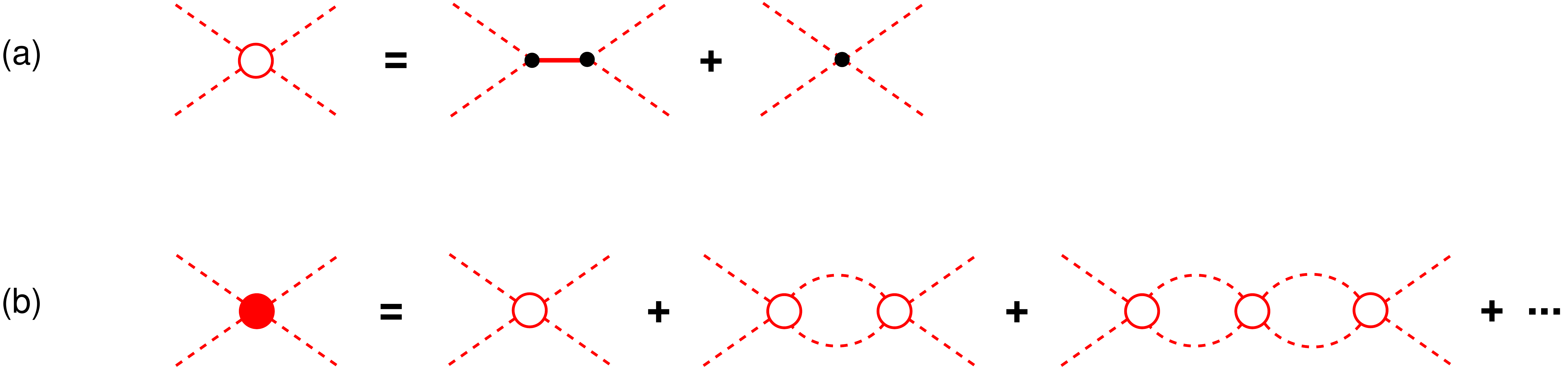}
\caption{(a) Tree-level two-body interaction kernel $\cK_{ab\rar cd}^{(J,I)}$ - 
a NLO $s$-channel resonance, added to a LO contact term.
(b) Structure of the unitarized scattering amplitude.}
\label{modA}
\end{figure}

The basic meson-meson intermediate interactions $ P^a P^b \rar P^c P^d$ 
are described  by  {\em kernels} $\cK_{ab|cd}^{(J,I)}$ 
and their simple dynamical structure is shown in Fig.\ref{modA},  as  LO four point terms, 
typical of chiral symmetry, supplemented by NLO resonance exchanges in  the $ s $-channel.
Just in the $ (J=0, I=0)$ channel two resonances, $S_1$ and $S_o$, are needed. 
In these diagrams, all {\em vertices} represent interactions derived from chiral lagrangians\cite{EGPR}.
Kernels are then functions depending on just masses and coupling constants. 
The mathematical structure of these functions is displayed in App.\ref{kernel}.
In the case of the $\phi$-meson, the kernel includes an effective coupling to the 
$(\rho \p + \p\p\p)$ channel, which accounts for about $15\% $ of its width. 
This effective interaction is discussed in App.(\ref{phi-propagator}) and yields eq.(\ref{k6}).

All other resonance terms in the kernels contain bare poles.
However, the evaluation of amplitudes involves the iteration of the basic kernels 
by means of two-meson  propagators, as in Fig.\ref{modA}(b).
The propagators, denoted by $ \Ob $, are discussed in App.\ref{omega} and, 
in principle, have both real and imaginary components.
The former contain divergent contributions and their regularization brings 
unknown parameters into the problem.   
This considerable nuisance is avoided by working in the  $ K $-matrix approximation,
whereby just the imaginary parts of the two-meson propagators are kept. 
This gives rise to the structure sketched within the square bracket of eq.(\ref{dyn.1}), 
where the terms $ (\mathrm{loop} \times \cK) $ are realized by 
the functions $ M_{ij}^{(J,I)} $ given in eqs.(\ref{z10}-\ref{z13}). 
The ressummation of the geometric series, indicated in Fig.\ref{modA}(b),
endows the  $ s $-channel  resonances with widths.
Thus among other structures, intermediate two-body amplitudes yield denominators $D^{(J,I)}$, 
which are akin to those of the form $D_{BW} =[s- m^2 + i\,m\, \Gamma]$ employed in BW functions.
These denominators, that correspond to the predictions of the model for the resonance line shapes, 
are given in App.\ref{decayT} and reproduced below. 
Explicit expressions  read
\bea
&& D_\rho 
= \lb \lp 1\sm M_{11}^{(1,1)}\rp  \,  \lp 1\sm M_{22}^{(1,1)}\rp  - M_{12}^{(1,1)}\, M_{21}^{(1,1)} \rb \;,
\label{un1}\\[4mm]
&& D_\phi 
= \lc 1\sm M^{(1,0)} \rc \,,
\label{un2}\\[4mm]
&& D_{a_0} 
= \lb \lp 1\sm M_{11}^{(0,1)} \rp \,  \lp 1\sm M_{22}^{(0,1)} \rp - M_{12}^{(0,1)}  M_{21}^{(0,1)}  \rb \,,
\label{un3} \\[4mm]
&& D_S 
= [1\sm M_{11}^{(0,0)}][1\sm M_{22}^{(0,0)} ][1 \sm M_{33}^{(0,0)}]
-  [1 \sm M_{11}^{(0,0)}]  M_{23}^{(0,0)} M_{32}^{(0,0)} 
\nn\\[2mm]
&& - [1 \sm M_{22}^{(0,0)}] M_{13}^{(0,0)} M_{31}^{(0,0)} 
- [1 \sm M_{33}^{(0,0)}] M_{12}^{(0,0)} M_{21}^{(0,0)} 
\nn\\[2mm]
&& -  M_{12}^{(0,0)}  M_{23}^{(0,0)} M_{31}^{(0,0)}  
- M_{21}^{(0,0)}  M_{32}^{(0,0)}  M_{13}^{(0,0)} \;,
\label{un4}
\eea
where the functions $M_{ij}^{(J,I)}$ read
\bea
&& M_{11}^{(1,1)} =  - \cK_{\p\p|\p\p}^{(1, 1)} \, [ \Ob_{\p\p}^P /2 ]  \;,
\hspace{10mm}
M_{12}^{(1,1)} =  -  \cK_{\p\p|KK}^{(1, 1)}  \, [\Ob_{KK}^P/2] \;,
\nn \\[2mm]
&& M_{21}^{(1,1)}  = - \cK_{\p\p|KK}^{(1, 1)} \, [\Ob_{\p\p}^P/2] \;,
\hspace{10mm}
M_{22}^{(1,1)} = -  \cK_{KK|KK}^{(1, 1)} \, [\Ob_{KK}^P /2] \;.
\label{un5}
\eea
\bea
&& M^{(1,0)} = - \cK_{KK|KK}^{(1, 0)} \, [\Ob_{KK}^P/2]\;. 
\label{un6}
\eea
\bea
 && M_{11}^{(0,1)} =  - \cK_{\p 8|\p 8}^{(0, 1)} \, [\Ob_{\p 8}^S /2] \;,
\hspace{10mm} 
M_{12}^{(0,1)} =  - \cK_{\p 8 |KK}^{(0, 1)} \, [\Ob_{KK}^S/2] \;,
\nn \\[2mm]
&& M_{21}^{(0,1)}  = - \cK_{\p 8|KK}^{(0, 1)} \, [\Ob_{\p8}^S /2] \;,  
\hspace{10mm}
M_{22}^{(0,1)} = - \cK_{KK|KK}^{(0, 1)} \, [\Ob_{KK}^S/2 ]\;. 
\label{un7}
\eea
\bea
&& M_{11}^{(0,0)} =  - \cK_{\p\p|\p\p}^{(0,0)} \, [\Ob_{\p\p}^S/2] \;, 
\hspace{10mm}
M_{12}^{(0,0)} =  - \cK_{\p\p|KK}^{(0,0)} \, [\Ob_{KK}^S/2] \;,
\nn\\[2mm]
&& M_{13}^{(0,0)} =  - \cK_{\p\p|88}^{(0,0)} \, [\Ob_{88}^S/2] \;,
\hspace{10mm}
M_{21}^{(0,0)} =  - \cK_{\p\p|KK}^{(0,0)} \, [\Ob_{\p\p}^S/2] \;,
\nn\\[2mm]
&& M_{22}^{(0,0)} =  - \cK_{KK|KK}^{(0,0)} \, [\Ob_{KK}^S/2] \;,
\hspace{10mm}
M_{23} ^{(0,0)}=  - \cK_{KK|88}^{(0,0)} \, [\Ob_{88}^S/2] \;,
\nn \\[2mm]
&& M_{31}^{(0,0)} =  - \cK_{\p\p|88}^{(0,0)} \, [\Ob_{\p\p}^S/2] \;,
\hspace{10mm}
M_{32}^{(0,0)} =  - \cK_{KK|88}^{(0,0)} \, [\Ob_{KK}^S/2] \;,
\nn \\[2mm]
&& M_{33}^{(0,0)} =  - \cK_{88|88}^{(0,0)} \, [\Ob_{88}^S/2] \;,
\label{un8}
\eea
with the $ \cK_{ab|cd}^{(J,I)} $ of App.\ref{kernel}, 
whereas the subscripts $8$ refer to the member of the $SU(3)$ octet with the quantum numbers of 
the $\eta$.
The factor $1/2$ in these expressions accounts for the symmetry of intermediate states
and it is also present in the functions $M_{11}^{(0,1)}$ and $M_{21}^{(0,1)}$
because one is using the symmetrized $\pi 8$ intermediate state 
given by eq.(\ref{su8}).

The imaginary propagators $\Ob$ of App.\ref{omega} are given by
\bea 
&& \Ob_{ab}^S = -\,\frac{i}{8\p} \; \frac{Q_{ab}}{\sqrt{s}} \;
\theta(s \sm (M_a \sp M_b)^2) \;,
\label{un9}\\[2mm]
&& \Ob_{aa}^P = -\,\frac{i}{6\p} \; \frac{Q_{aa}^3}{\sqrt{s}} \;
\theta(s \sm4\, M_a^2) \;,
\label{un10}\\[2mm]
&& Q_{ab} = \frac{1}{2} \, \sqrt{s - 2\,(M_a^2 + M_b^2) + (M_a^2 - M_b^ 2)^2/s} \;,
\label{un11}
\eea 
$\theta$ being the Heaviside step function.

The dynamical meaning of the functions $\Ob_{ab}^J$ and $M_{ab}^{(J,I)}$
is indicated in Fig.\ref{modA}(b).
The former represents the two-body propagator for mesons $a$ and $b$ with 
angular momentum $J$,  indicated by the dashed lines between two successive empty blobs,
whereas the latter encompasses a blob and a two-body propagator.
The functions $M_{ab}^{(J,I)}$ correspond to the paces of the the various geometric series
entangled by the coupling of intermediate channels.

\subsection{$K \Kb $ scattering amplitude}

The $ K \Kb $ scattering amplitude, which is a prediction of the model, 
is derived in App.\ref{scatt} and is written in terms of 
the denominators $ D^{(J,I)} $ as 
\bea
&& A_{KK|KK}^{(1,1)} = \frac{1}{D_\rho (m_{12}^2)}
\lb M_{21}^{(1,1)} \, \cK_{\p\p|KK}^{(1,1)} + \lp 1 \sm M_{11}^{(1,1)}\rp \, \cK_{KK|KK}^{(1,1)}  \rb \;,
\label{sca1}\\[4mm] 
&& A_{KK|KK}^{(1,0)} =  \frac{1}{D_\phi (m_{12}^2)}\;  \cK_{KK|KK}^{(1,0)} \;,
\label{sca2}\\[4mm]
&& A_{KK|KK}^{(0,1)} = \frac{ 1}{D_{a_0} (m_{12}^2)}
\lb M_{21}^{(0,1)} \,  \cK_{\p 8|KK}^{(0,1)} + \lp 1 \sm M_{11}^{(0,1)} \rp \, \cK_{ KK|KK}^{(0,1)}  \rb 
\label{sca3}\\[4mm] 
&& A_{KK|KK}^{(0,0)} = \frac{1}{D_S(m_{12}^2)}
\lc \lb M_{21}^{(0,0)} \lp 1\sm M_{33}^{(0,0)} \rp \sp M_{23}^{(0,0)} M_{31}^{(0,0)}\rb  \, \cK_{\p\p|KK}^{(0,0)}
\right.
\nn\\[2mm]
& & \left.  
\, + \lb \lp 1\sm M_{11}^{(0,0)} \rp \lp1\sm M_{33}^{(0,0)} \rp \sm M_{13}^{(0,0)} M_{31}^{(0,0)}\rb \, \cK_{KK|KK}^{(0,0)} 
\right.
\nn\\[2mm]
&& \left.  
\, + \lb M_{23}^{(0,0)} \lp 1\sm M_{11}^{(0,0)} \rp \sp M_{13}^{(0,0)} M_{21}^{(0,0)}\rb  \, \cK_{88|KK}^{(0,0)} \rc \;.
\label{sca4}
\eea

\subsection{decay amplitude}
\label{secsec}

The decay amplitude for the process $D^+ \rar K^- \, K^+ \, K^+ $, given by eq.(\ref{mmm.2}),
has the general structure 
\bea
T  &\!=\!& T_{NR} + \lb T^{(1,1)}  + T^{(1,0)} +  T^{(0,1)} + T^{(0,0)} + (2\lrar 3) \rb \;,
\label{dec1}
\eea
where $ T_{NR} $ is the non-resonant contribution from diagrams (1A+1B) of Fig.\ref{modT}
and the $ T^{(J,I)} $
are the resonant contributions from diagrams (2A+2B+3A+3B+4A+4B), 
in the various spin and isospin channels.

Owing to chiral symmetry, all amplitudes are proportional to $ M_K^2 $,
included in a common factor
\bea
C  = \lc \lb  \frac{G_F}{\rtw} \, \sin^2\theta_C \rb \; 
\frac{2 F_D}{F} \, \frac{M_K^2}{(M_D^2-M_K^2)} \rc \,,
\label{dec1a}
\eea
where $ F $ is the $ SU(3) $ pseudoscalar decay constant.
Using the kinematical variables $ m_{ij}^2 = (p_i \sp p_j)^2 $,
the non-resonant contribution is the real polynomial
\bea
&& T_{NR}  =  C\, \lc \lb (m_{12}^2-M_K^2) + (m_{13}^2-M_K^2) \rb \rc \;,
\label{dec2}
\eea
corresponding to a proper three-body interaction.
The amplitudes $T^{(J,I)}$ read
\bea
&& T^{(1,1)}  = -\, \frac{1}{4}\, \lb \Gb_{KK}^{(1,1)}  - \G_{c|KK}^{(1,1)} \rb  \, (m_{13}^2 \sm m_{23}^2)\,,
\label{dec3} \\[2mm]
&& \Gb_{KK}^{(1,1)} = \frac{1}{D_\rho (m_{12}^2)}
\lb M_{21}^{(1,1)}  \, \G_{(0)\, \p\p}^{(1,1)} + \lp 1 \sm M_{11}^{(1,1)} \rp \, \G_{(0)\, KK}^{(1,1)}  \rb \,,
\label{dec4} \\[4mm] 
&& T^{(1,0)}  = -\,  \frac{1}{4}\,
\lb \Gb_{KK}^{(1,0)}  - \G_{c|KK}^{(1,0)}  \rb  \, (m_{13}^2 \sm m_{23}^2) \,,
\label{dec5} \\[2mm]
&& \Gb_{KK}^{(1,0)} 
= \frac{1}{D_\phi (m_{12}^2)}\; \G_{(0)\, KK}^{(1,0)} \,,
\label{dec6} \\[4mm]
&& T^{(0,1)}  =  -\, \frac{1}{2}\, \lb \Gb_{KK}^{(0,1)}  - \G_{c|KK}^{(0,1)} \rb  \;,
\label{dec7} \\[2mm]
&& \Gb_{KK}^{(0,1)} = \frac{1}{D_{a_0} (m_{12}^2)}
\lb M_{21}^{(0,1)}  \, \G_{(0)\, \p 8}^{(0,1)} + \lp 1 \sm M_{11}^{(0,1)} \rp \, \G_{(0)\, KK}^{(0,1)}  \rb \,,
\label{dec8} \\[4mm] 
&& T^{(0,0)}  =  -\, \frac{1}{2}\, \lb \Gb_{KK}^{(0,0)}  - \G_{c|KK}^{(0,0)} \rb  \,,
\label{dec9}\\[2mm]
&& \Gb_{KK}^{(0,0)} = \frac{1}{D_S(m_{12}^2) }
\lc \lb M_{21}^{(0,0)} \lp 1\sm M_{33}^{(0,0)}\rp \sp M_{23}^{(0,0)} M_{31}^{(0,0)}\rb \, \G_{(0)\,\p\p}^{(0,0)} 
\right.
\nn \\[2mm]
&& \left. + \lb \lp 1\sm M_{11}^{(0,0)}\rp \lp 1\sm M_{33}^{(0,0)}\rp  
\sm M_{13}^{(0,0)} M_{31}^{(0,0)}\rb  \, \G_{(0)\,KK}^{(0,0)} 
\right.
\nn\\[2mm]
&&  \left. +\, 
\lb  M_{23}^{(0,0)} \lp 1\sm M_{11}^{(0,0)} \rp  \sp M_{13}^{(0,0)} M_{21}^{(0,0)} \rb  \, \G_{(0)\,88}^{(0,0)} \rc \;,
\label{dec10}
\eea
where the various functions $\G^{(J,I)} $,  given in App.\ref{treedec}, are linear in the coefficient $C$.
The dynamical meaning of the functions $\G_{(0)ab}^{(J,I)} $  can be inferred from Fig.\ref{modT}(b).
They correspond to the tree diagrams (1A+1B) and (3A+3B) with the indices $(1,2) \rar (a,b)$
and represent the amplitude for the production of pseudoscalar mesons $P^a P^b K^+$
by a $W^+$.

Comparing results (\ref{dec3}-\ref{dec10}) and  (\ref{sca1}-\ref{sca4}), it is easy to see that 
the decay amplitudes $T^{(J,I)} $ and the scattering amplitudes $A^{(J,I)} $ are quite different objects,
since the former include the weak interaction, which is encoded into the decay vertices $ \Gb_{KK}^{(J,I)}$.
Nevertheless, both $A_{KK|KK}^{(J,I)} $  and $ \Gb_{KK}^{J,I)} $ share the same denominators $D^{(J,I)}$. 
The amplitude $T$, given by eq.(\ref{dec1}) is our guess function, to be used in fits to data.
As it is a blend of spin and isospin channels, attempts to compare it directly to the $A^{(J,I)}$ are 
meaningless.

\subsection{free parameters}

The free parameters of our function $T$  derive from the basic lagrangian adopted \cite{EGPR}
and consist basically of masses and coupling constants.
The former include $m_\rho, m_\phi, m_{a0}, m_{S1}, m_{So}$, whereas the latter involve 
$F$, the pseudoscalar decay constant, $G_V$, the coupling constant of vector mesons to pseudoscalars, 
an angle $\theta$, associated with $\omega\!-\!\phi$ mixing,  
$c_d, c_m $, describing the couplings of both $a_0$ and $S_o$ to pseudoscalars, and 
$\ct_d, \ct_m$, implementing the couplings of $S_1$ to pseudoscalars.
These lagrangian parameters first enter the  guess function 
through the functions $\G_{(0)\,ab}^{(J,I)}$  and $\cK_{ab|cd}^{(J,I)}$ 
in apps. \ref{treedec} and \ref{kernel}.

In the strict framework of chiral perturbation theory, the values of the lagrangian parameters 
are extracted by comparing results from field theoretical calculations performed to 
a given order to  observables.
As the former involve divergent loops, they are affected by renormalization and 
values quoted in the literature depend on renormalization scales.
This kind of procedure is theoretically consistent and yields a precise 
description of low-energy phenomena.

In the case of heavy meson decays, this level of precision cannot be reached.
The main reason is that the problem involves necessarily a wide range of energies,
both below and above resonance poles, where perturbation does not apply 
and non-perturbative techniques are needed.  
An instance is the resummation of the infinite series of diagrams indicated in Fig.\ref{modA},
required by unitarization, which yields the denominators $D^{(J,I)}$ discussed in sect.\ref{denominators}.
Therefore, in decay analyses, the free parameters
do not have the same meaning as their low-energy counterparts,
since they are designed to be used into a mathematical structure which is different from ChPT.
The former correspond to  effective parameters describing the physics
within the energy ranges defined by Dalitz plots and
should not be expected to have the same values as the latter.

\section{a toy example: decay $\times$ scattering amplitudes}
 
The Triple-M is aimed at predicting scattering amplitudes by 
using parameters obtained from fits to decay data. 
Even in the want of such fitted parameters at present, we explore the features of the lagrangian 
by using those suited to problems at low-energies,  which are:
$[ m_\rho, m_\phi, m_{a0}, m_{So}] = [0.776, 1.019, 0.960, 0.980]\,$GeV\cite{PDG},
$F=0.093\,$GeV, 
$[G_V, c_d, c_m, \ct_d, \ct_m]=[0.067, 0.032, 0.042, 0.018, 0.025]\,$GeV\cite{EGPR},
whereas the partial width $\Gamma_{\phi\rar K\Kb} \sim 3.54 \,$MeV\cite{PDG} yields $\sin\theta=0.605$. 
In the large $N_C$ limit, $m_{S1}=m_{So} $\cite{EGPR} but, in order to 
perform the toy calculations, we choose  $m_{S1}  = 1.370\,$GeV\cite{PDG}.
The discussion presented in the sequence  makes it clear
that there is no simple relation between the decay amplitude $T$ and the scattering 
amplitudes $A^{(J,I)}$.

The non-resonant contribution to the decay amplitude, eq.(\ref{dec2}), 
corresponds to a genuine three-body interaction predicted by chiral symmetry. 
Nevertheless, in order to assess  its relative importance,  it is convenient to project 
it into  the $S$- and $P$-waves suited to the other terms.
Therefore, we rewrite it as
\bea
&& T_{NR}  =  \lb \frac{C}{4} \, (M^2 - M_K^2+ m_{12}^2)
+ \frac{C}{4} \,  (m_{13}^2-m_{23}^2) + (2\lrar 3) \rb \;,
\label{dXs.1}
\eea
so that the  amplitude (\ref{dec1}) can then be expressed as 
\bea
T  &\!=\!& \lb T^S + T^P + (2 \lrar 3) \rb \;,
\label{dXs.2}\\[2mm]
T^S   &\!=\!& \lb \frac{C}{4} \, (M_D^2 - M_K^2+ m_{12}^2) + T^{(0,1)}  + T^{(0,0)} \rb \;,
\label{dXs.3}\\[2mm]
T^P   &\!=\!& \lb \frac{C}{4} \,  (m_{13}^2-m_{23}^2)  + T^{(1,1)}  + T^{(1,0)} \rb \;.
\label{dXs.4}
\eea

\begin{figure}[t] 
\includegraphics[width=0.48\columnwidth,angle=0]{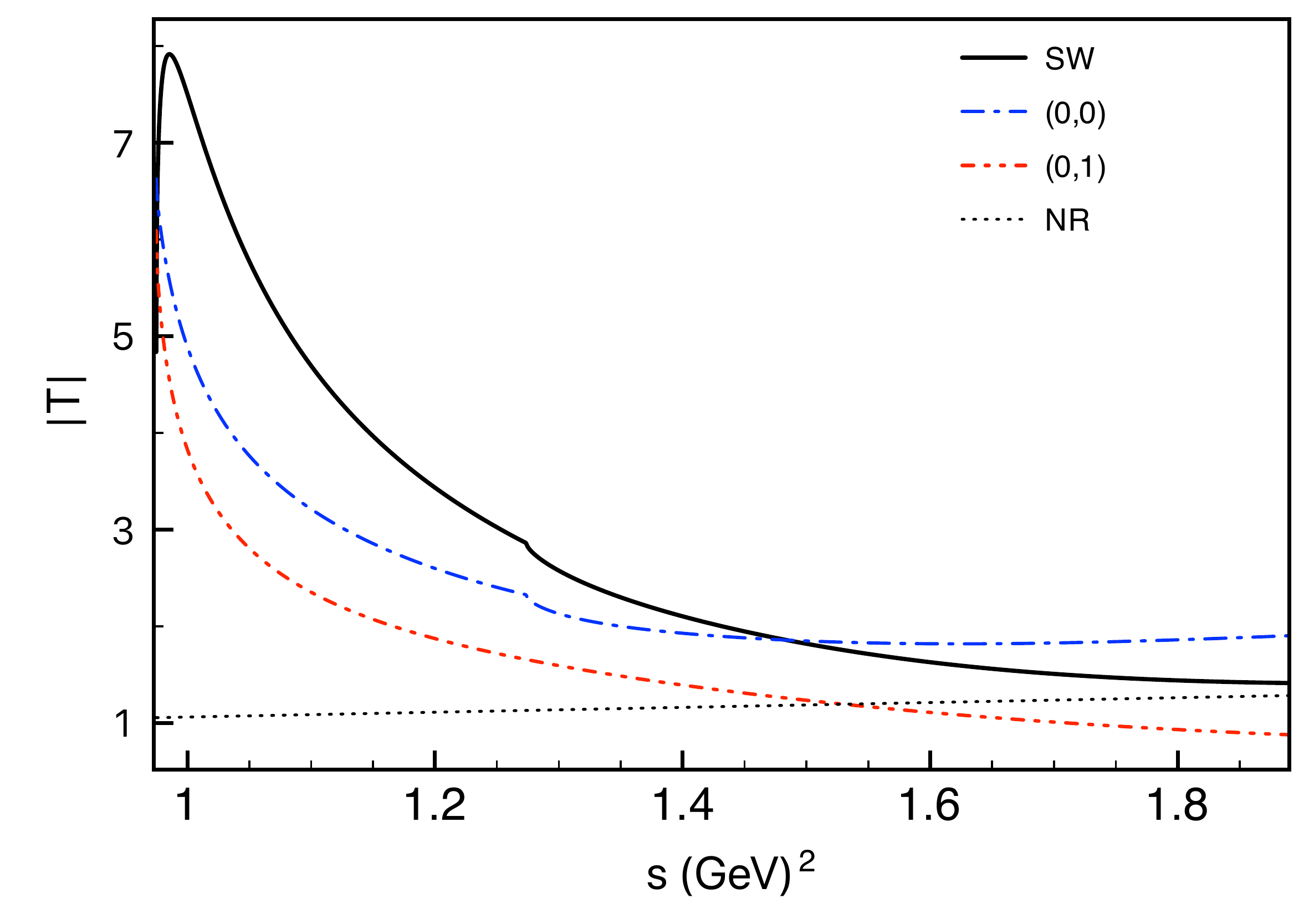}
\includegraphics[width=0.48\columnwidth,angle=0]{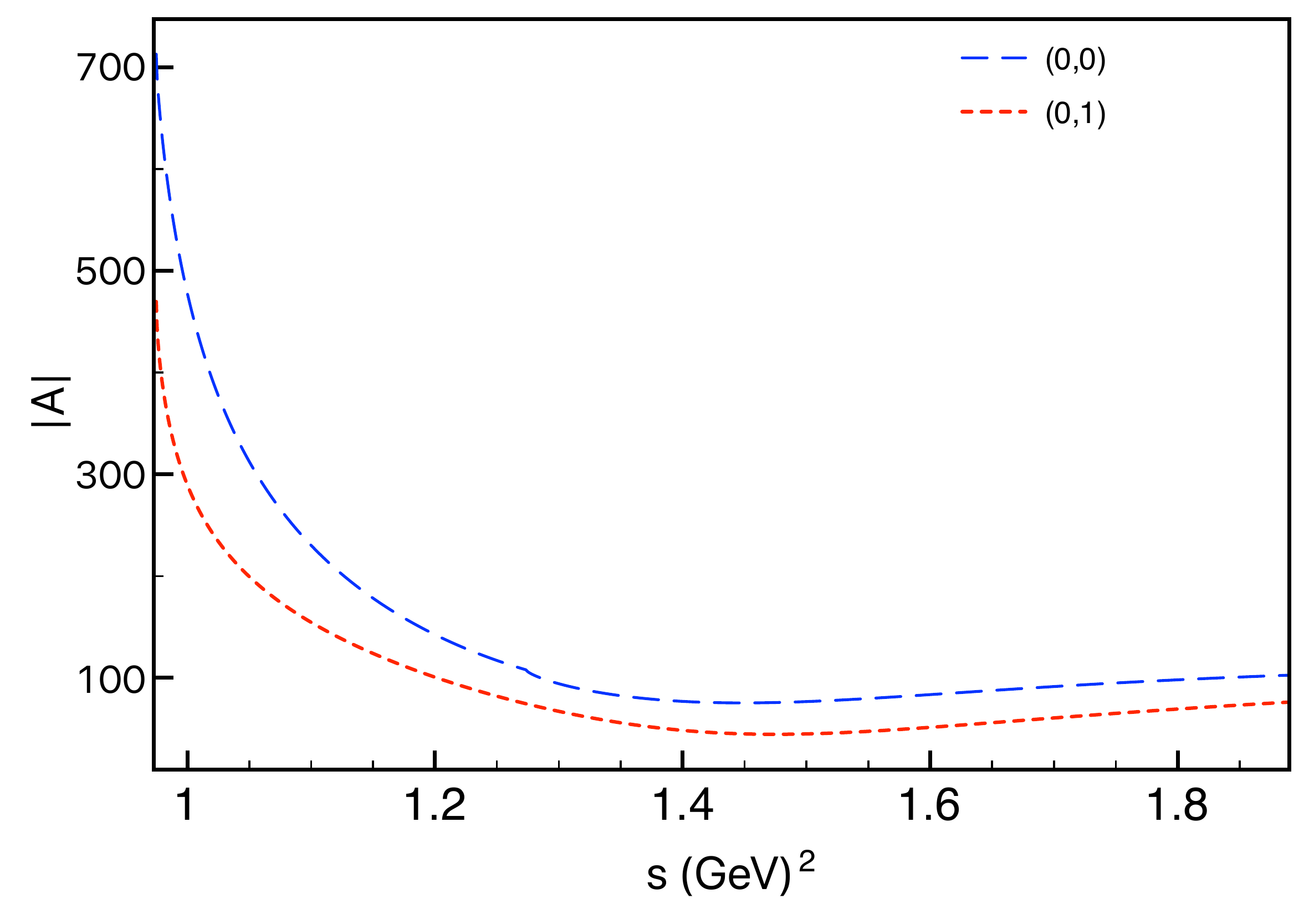}
\\
\includegraphics[width=.7\columnwidth,angle=0]{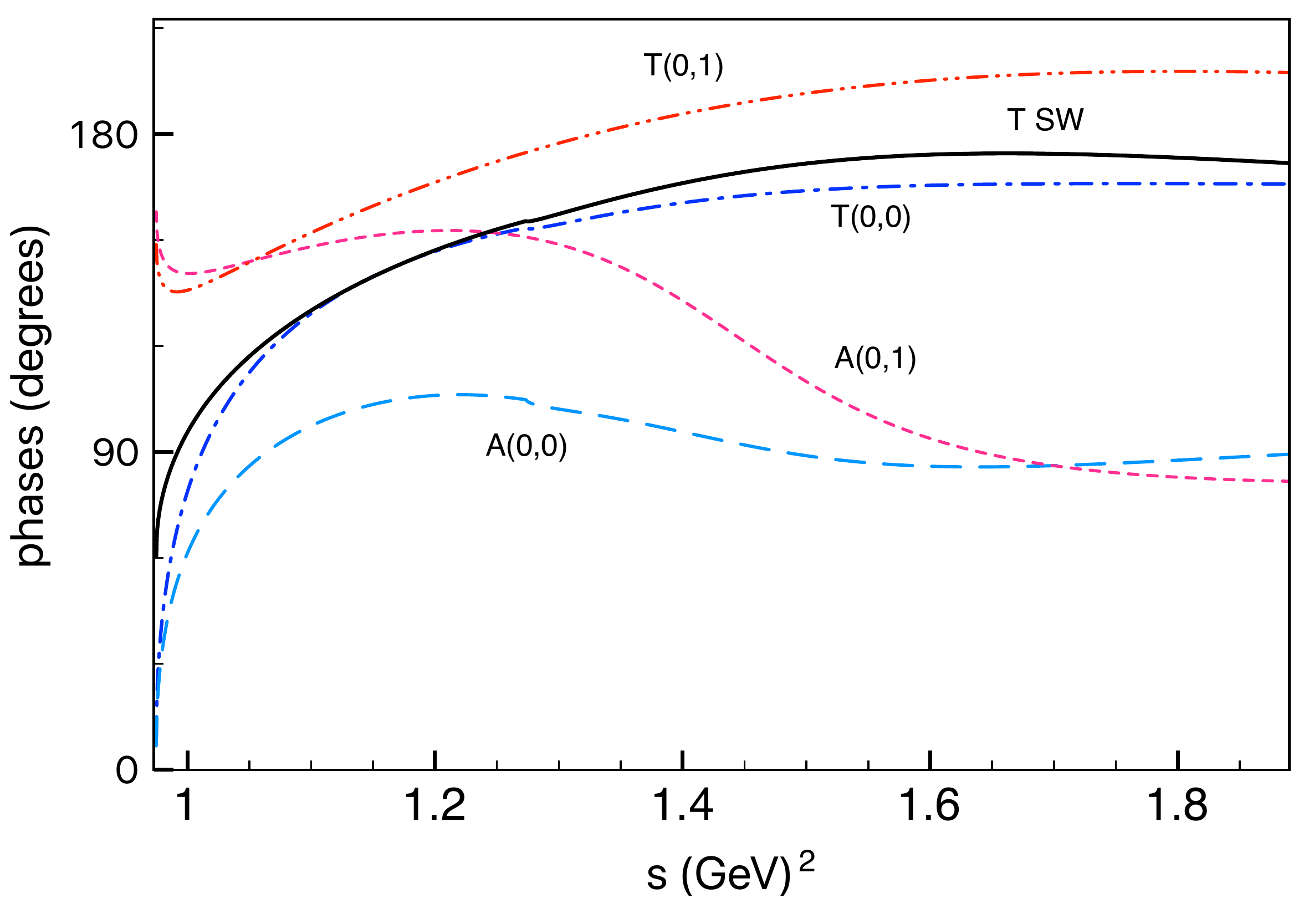}
\\
\caption{$S$-wave sector -
top left: the continuous black curve (SW) is the modulus of the decay amplitude 
$T^S$, eq.(\ref{dXs.3}), in arbitrary units, whereas other curves are moduli of
 partial contributions;
top right: moduli of the $K\Kb$ scatterig amplitudes $A^{(0,1)}$,
red curve, and $A^{(0,0)}$, blue curve; 
bottom: the continuous black curve (SW) is the phase of the decay amplitude 
$T^S$, eq.(\ref{dXs.3}), and other continuous curves are phases of partial contributions;
the dashed curves represent the phases of the $K\Kb$ scatterig amplitudes $A^{(0,1)}$ (red)
and $A^{(0,0)}$ (blue).}
\label{Swave}
\end{figure}

\begin{figure}[t] 
\includegraphics[width=0.48\columnwidth,angle=0]{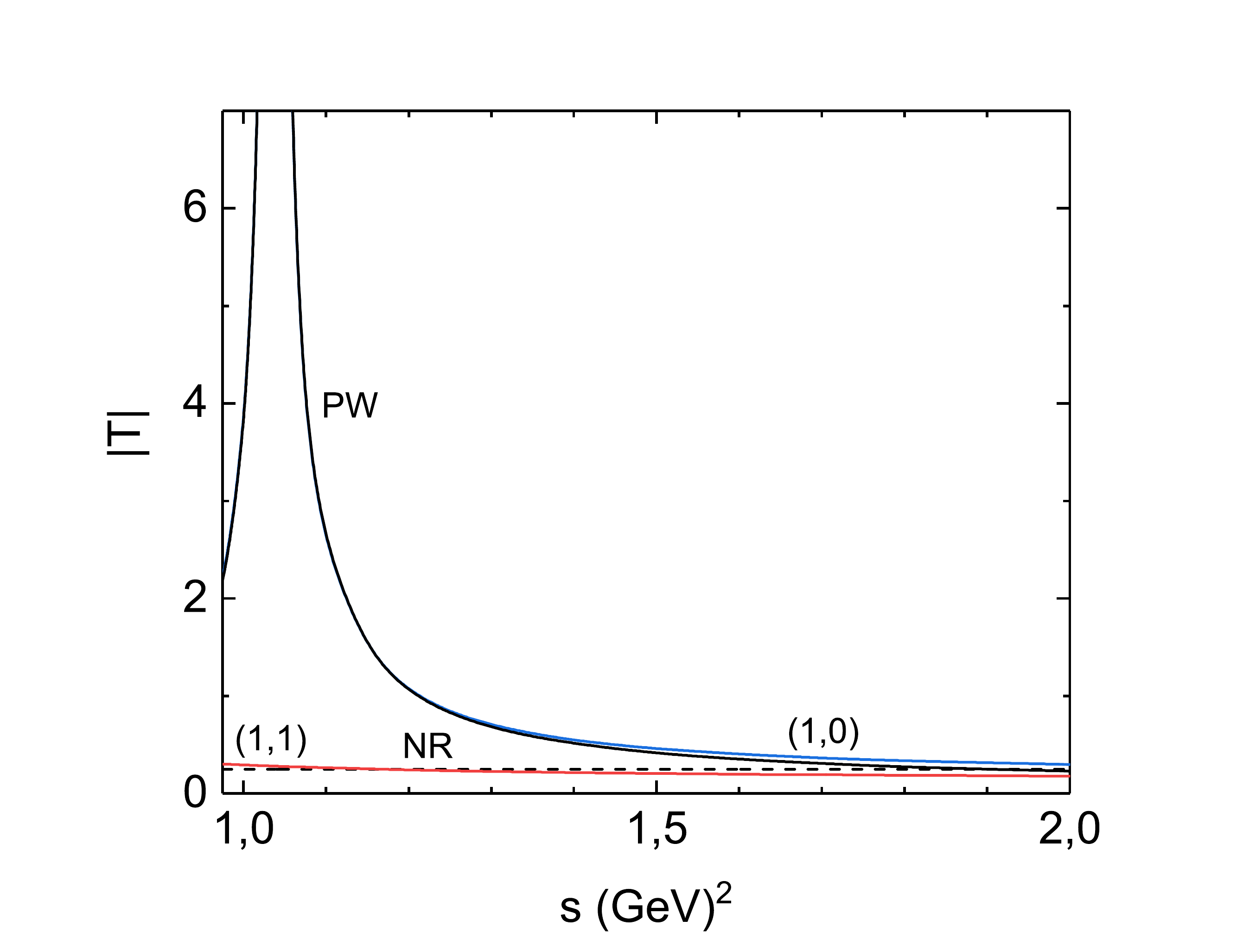}
\includegraphics[width=0.48\columnwidth,angle=0]{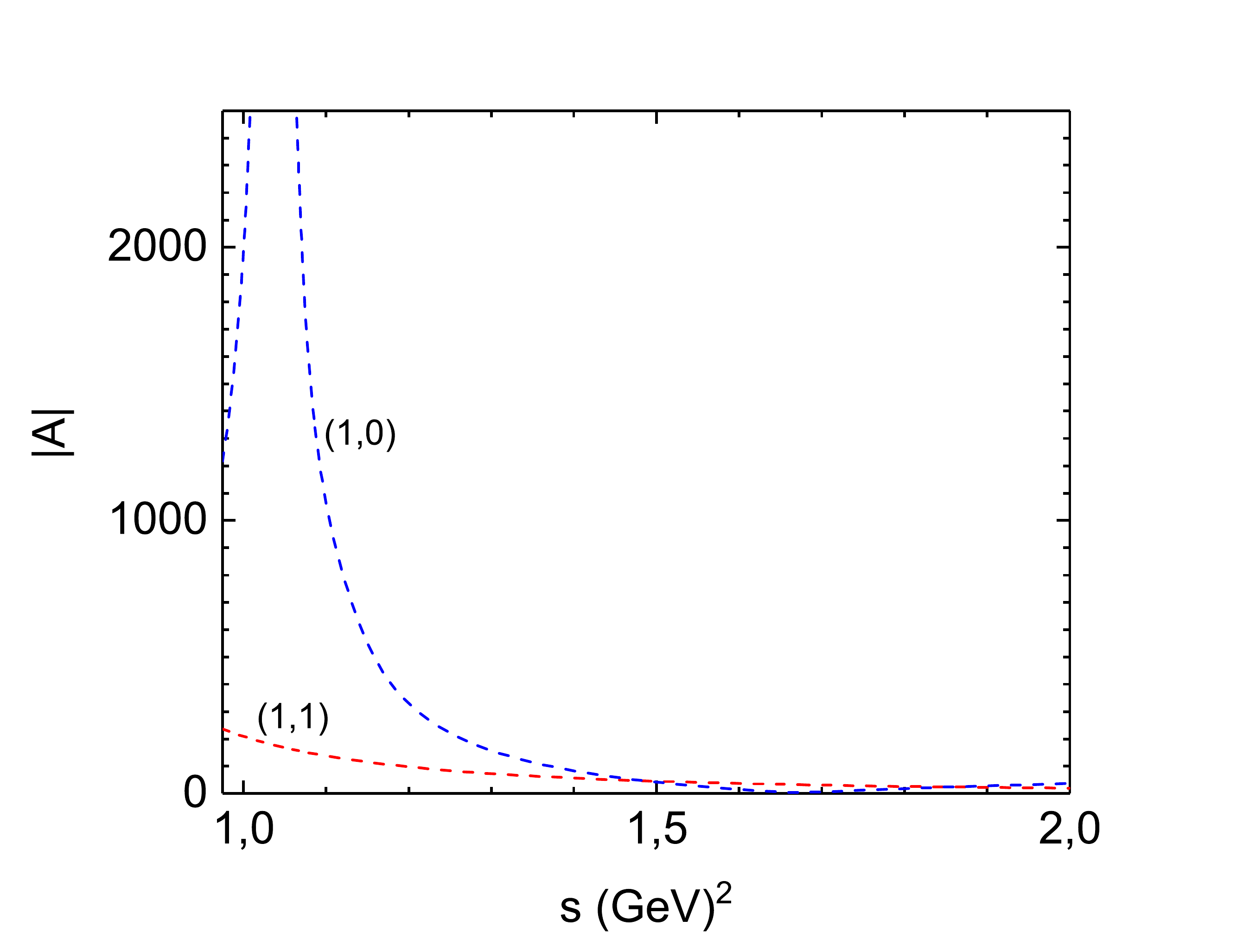}
\\
\includegraphics[width=.7\columnwidth,angle=0]{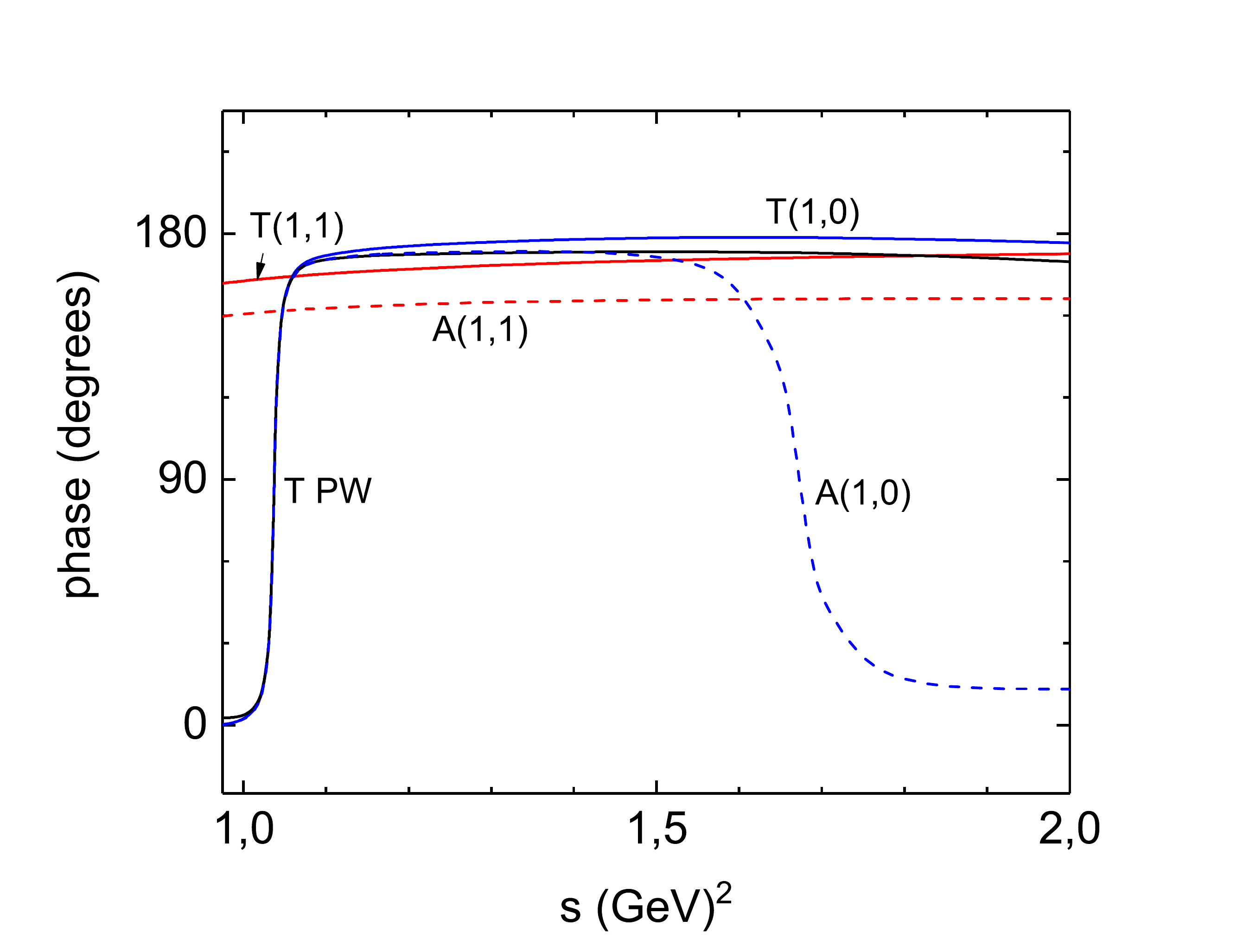}
\\
\caption{$ P $-wave sector -
top left: the continuous black curve (SW) is the modulus of the decay amplitude 
$T^P$, eq.(\ref{dXs.4}), in arbitrary units, whereas other curves are moduli of partial
 contributions;
top right: muduli of the $K\Kb$ scatterig amplitudes $A^{(1,1)}$,
red curve, and $A^{(1,0)}$, blue curve; 
bottom: the continuous black curve (SW) is the phase of the decay amplitude 
$T^S$, eq.(\ref{dXs.3}), and other continuous curves are phases of partial contributions;
the dashed curves represent the phases of the $K\Kb$ scatterig amplitudes $A^{(1,1)}$ (red)
and $A^{(1,0)}$ (blue).}
\label{Pwave}
\end{figure}

In the sequence, we discuss some aspects of this relationship, using the low-energy 
parameters of Ref.\cite{EGPR}, as if they could explain decay data.
In Figs.\ref{Swave} and \ref{Pwave}, we show the moduli and phases of 
the $S$- and $P$-wave decay amplitudes $T^S$, eq.(\ref{dXs.3}) and $T^P$, eq.(\ref{dXs.4}),
together with the moduli and phases of the corresponding $ K\Kb $ scattering 
amplitudes $A^{(J,I)}$.
These figures illustrate the usefulness of the lagrangian approach.
Without it, one would be able to determine just the full decay amplitudes $T^S$ and $ T^P$, 
represented by the continuous black curves in the figures, and 
would not have access to partial contributions in different isospin channels.
Moreover, it is also clear that one cannot guess the form of the $K\Kb $ scattering amplitudes
$A^{(J,I)} $, represented by the red and blue dotted lines, from the 
decay components $T^S$ and $T^P$.

\begin{figure}[t] 
\hspace*{-0.4cm}
\includegraphics[width=0.39\columnwidth,angle=0]{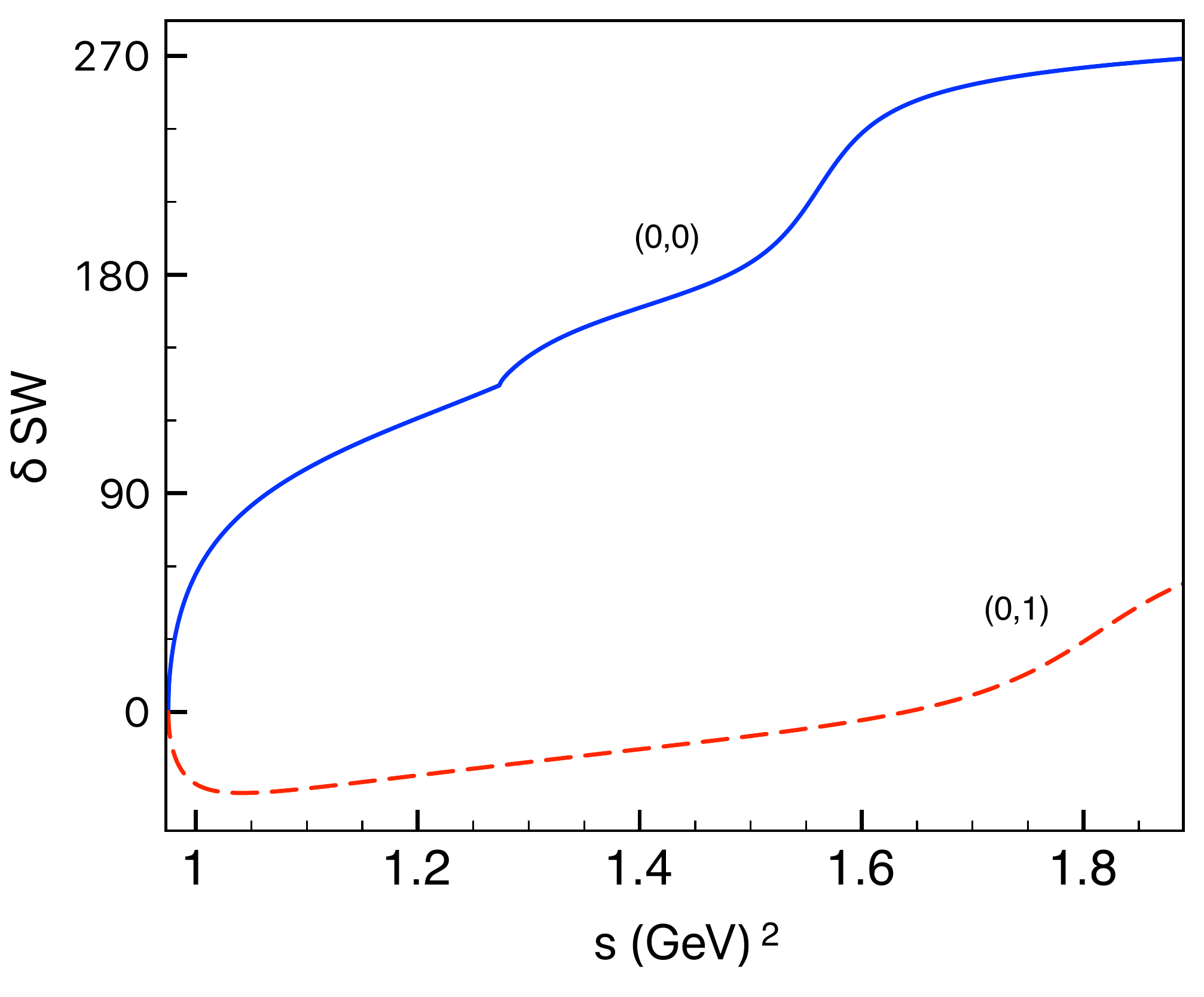}
\hspace*{1.4cm}
\includegraphics[width=0.39\columnwidth,angle=0]{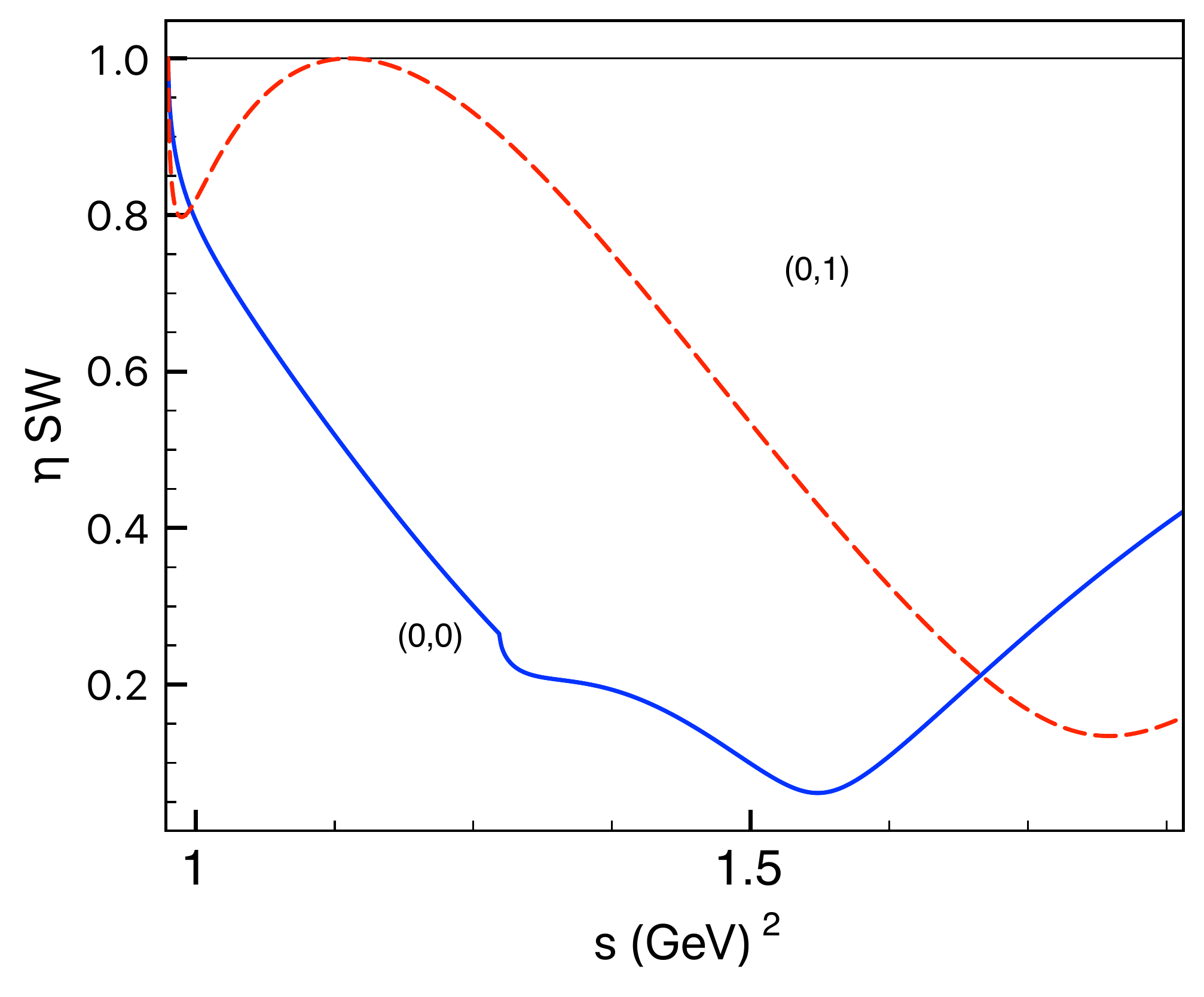}
\hspace*{0.2cm}
\\
\includegraphics[width=0.48\columnwidth,angle=0]{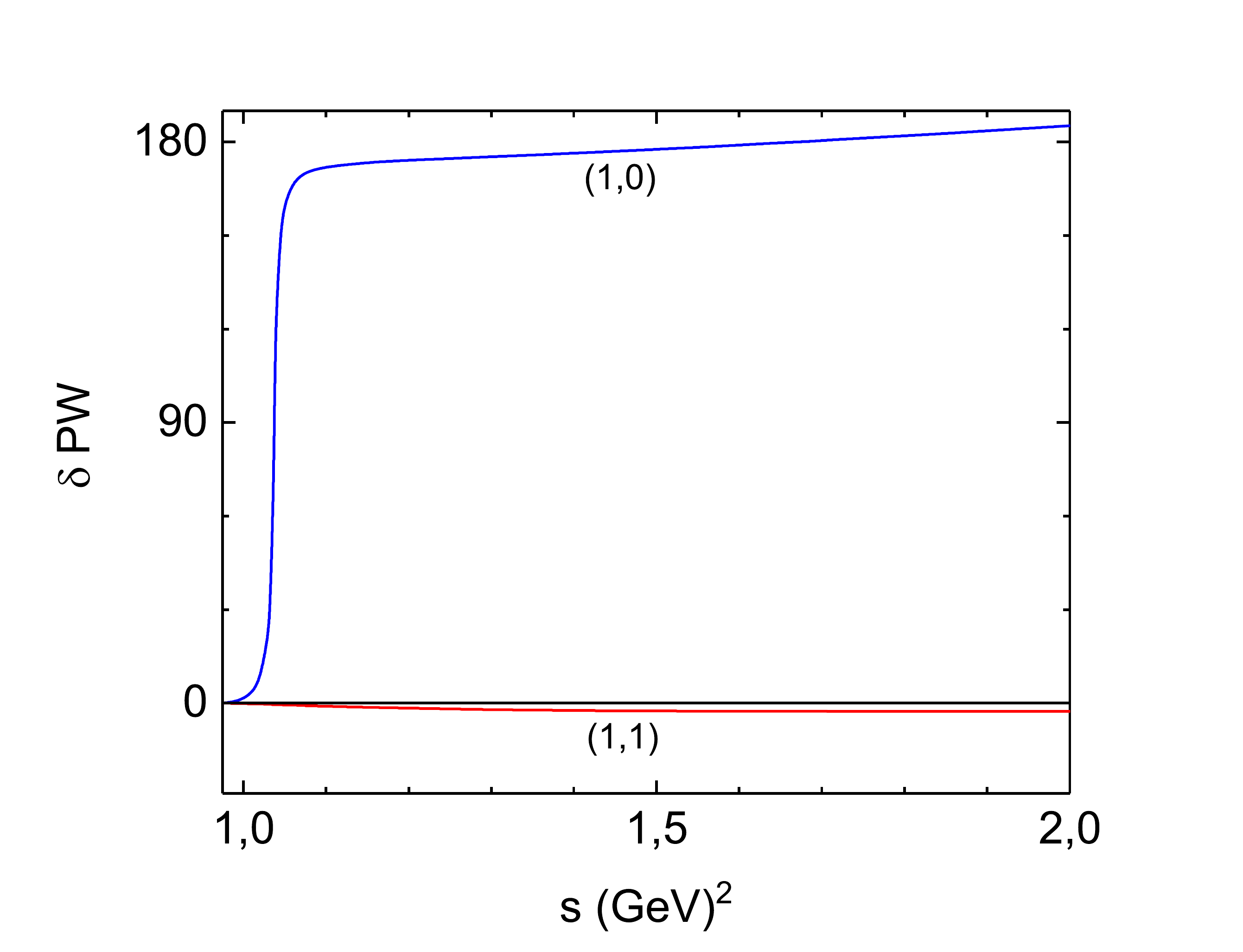}
\includegraphics[width=0.48\columnwidth,angle=0]{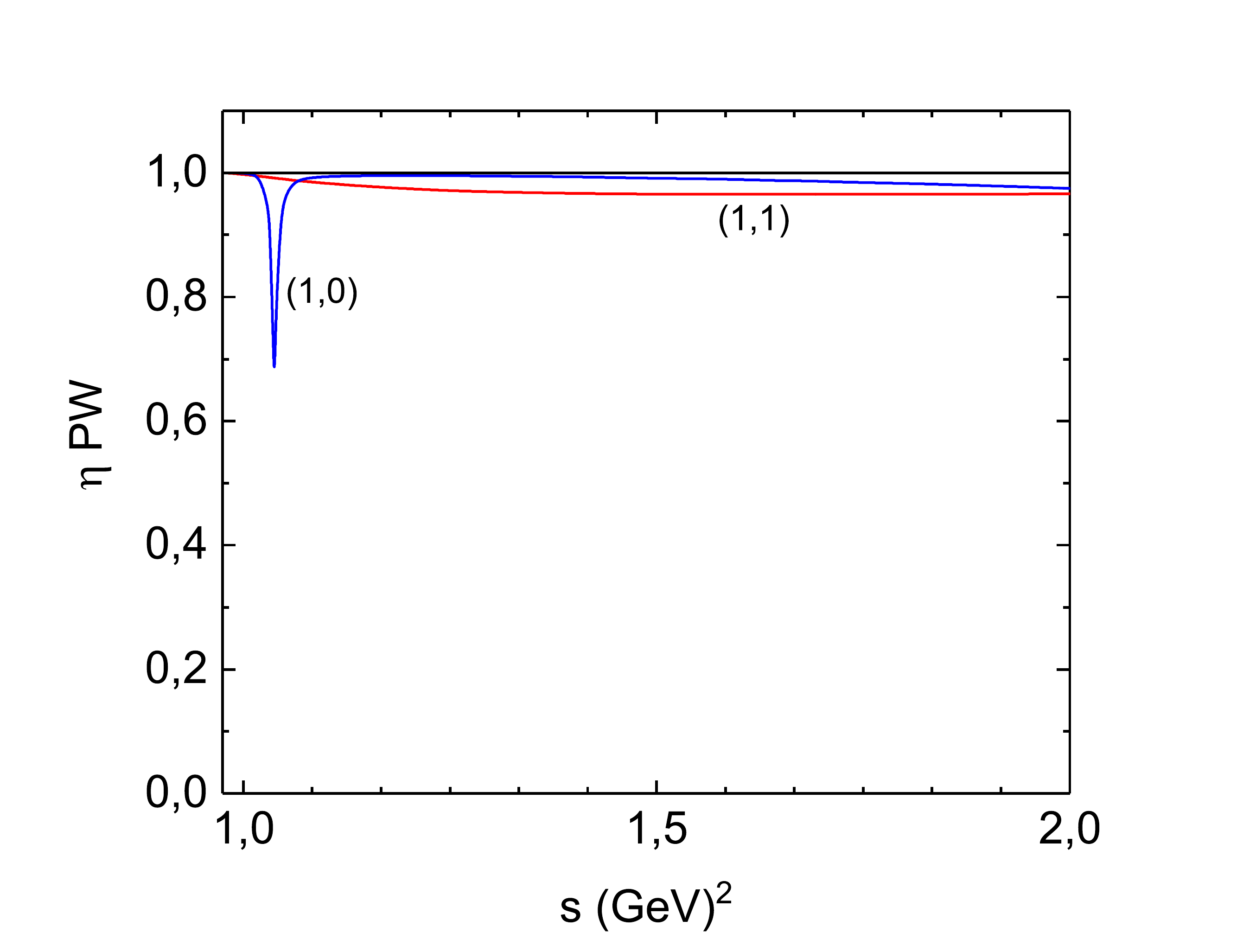}
\caption{Phase shifts $\d$ and inelasticity parameter $\eta $ for $K\Kb$ scattering -
top: $S$-waves; bottom: $P$-waves;
blue and red curves correspond respectively to isospin I=0 and I=1.}
\label{deltaeta}
\end{figure}

In Fig.\ref{deltaeta} we present the phase shifts and inelasticity parameters
associated with the scattering amplitudes $A^{(J,I)}$.
It important to stress that these figures correspond just to an exercise, since they
are based on low-energy parameters.
Nevertheless, they are instructive in showing the importance of the coupled channel structure,
which is responsible for the inelasticities displayed.
In the case of the I=1 $P$-wave, this related with the $\phi \rar \p\p\p $ channel,
as discussed in App.\ref{phi-propagator}.
In all cases, the bound $\eta \leq 1$ is satisfied.

The Multi-Meson-Model we consider here yields scattering amplitudes 
involving dynamical features such:
i) a chiral contact interaction in the two-body kernel,  indicated in Fig.\ref{modA};
ii) the use of two resonances in the $(J=0,I=0)$ channel, preserving unitarity;
iii) inclusion of coupled channels.
In App.\ref{case} we discuss their piecemeal relevance, in the case of $A^{(0,0)}$.

\section{Summary}
We propose a multi-meson-model (Triple-M) to describe the $\dkkk$ decay, as a tool to extract 
information about $K\bar{K}$ scattering amplitudes.
 We depart from the  dominance of the annihilation weak 
topology, which allows one to describe the whole decay process within the $SU(3)$ 
chiral symmetry framework.
The non-resonant component is a proper three-body interaction that goes beyond the (2+1)
 approximation and is given by chiral symmetry as a real polynomium.
Primary vertices describing the direct production of mesons and of lowest 
SU(3) resonances, in $S$- and $P$-waves, with isospin $0$ and $1$,
are dressed by FSIs involving coupled channels. 
The $K\bar{K}$ scattering amplitudes for each of the $(J,I)$ considered
are derived from the ChPTR Lagrangian\cite{EGPR},
unitarized by ressummation techniques in the $K$-matrix 
approximation, in which particle propagators were kept on-shell,
and include coupled-channels.
They are the only source of imaginary terms in the decay amplitude
and fix the relative phase between $S$- and $P$-waves in Triple-M.
This  represents an important improvement over the isobar model, where this phase is a fitting parameter.

The fitting parameters in the Triple-M  are resonance masses and coupling constants, which 
have a rather transparent physical meaning. Although they entered the Triple-M 
through the ChPTR Lagrangian, their meanings change so as to
accommodate non-perturbation effects of meson-meson interactions. 
To obtain realistic values for these parameters, they should be extracted from a Triple-M fit
to data. 
As a lesser alternative, here we employ the low-energy parameters\cite{EGPR} values as if 
they resulted from data. 
%
\begin{figure}[h] 
\includegraphics[width=0.45\columnwidth,angle=0]{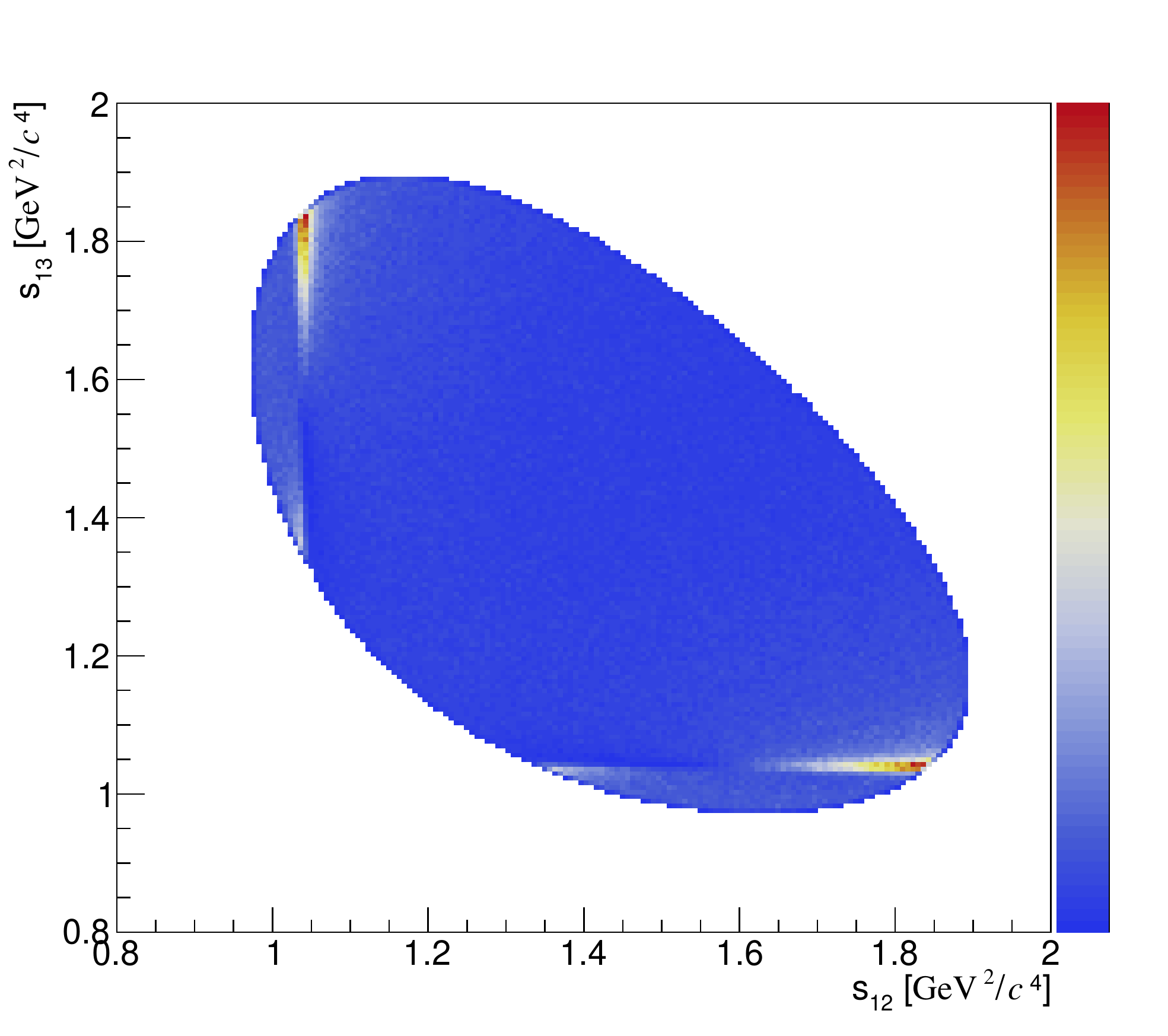}
\caption{Toy Dalitz plot for Triple-M in $D^+ \rar K^-\,K^+\, K^+$ decay with arbitrary normalization. }
\label{dalitz}
\end{figure}
In Fig.\ref{dalitz}, we show a toy Monte-Carlo Dalitz plot based on the Triple-M,
where it is possible to  see a  destructive interference between the $S$- and $P$-waves
on the low-energy sector of the $\phi(1020)$.  
One of the $\phi(1020)$ lobes is depleted with respect to the other, resulting in a peak and a dip,
a behaviour similar to that observed in LHCb preliminary data\cite{LHCbkkk}. 

In our one-dimensional toy studies, Figs.\ref{Swave}-\ref{Pwave}, we show that the Triple-M can 
track the hidden isospin signatures 
of  two-body  interactions in three-body data, 
allowing one to disentangle the relative contributions of  
resonances $a_0(980)$  and $f_0(980)$. 
By comparing results for the three-body amplitudes 
$T^J$ and the scattering amplitudes $A^{(J,I)}$, it becomes clear that  even though the 
later are present in the former, they cannot be extracted directly. 
However, with a model departing from a Lagrangian that includes 
a full two-body coupled channel dynamics, such as our Triple-M, fits to decay data 
can give rise to predictions for the $K\bar{K}$  scattering amplitudes $A^{(J,I)}$.

\section*{ACKNOWLEDGMENTS}
This work was supported by Conselho Nacional de Desenvolvimento Cient\'{i}fico e Tecnol\'{o}gico (CNPq).

\appendix

\section{kinematics}
\label{kin}

Momenta are defined by $ D(P) \rar  K^-(p_1)\, K^+(p_2) \, K^+(p_3) $, with
$ P=p_1 + p_2 + p_3 $.
The invariant masses read
\bea
&& m_{12}^2= (p_1 \sp p_2)^2 = (P \sm p_3)^2  \;,
\label{k.1}\\[2mm]
&& 
m_{13}^2= (p_1 \sp p_3)^2 = (P \sm p_2)^2 \;,
\label{k.2}\\[2mm]
&& 
m_{23}^2= (p_2 \sp p_3)^2 = (P \sm p_1)^2 \;,
\label{k.3}
\eea
and satisfy the constraint
\beq
M^2 = m_{12}^2 + m_{13}^2 + m_{23}^2 - m_1^2 - m_2^2 - m_3^2 \;.
\label{k.4}
\eeq
Their values are also limited by the boundaries of the Dalitz plot, by
\bea
&& (m_1 \sp m_2)^2 \leq m_{12}^2 \leq (M-m_3)^2 \;,
\label{k.5} \\[2mm]
&& (m_1 \sp m_3)^2 \leq m_{13}^2 \leq (M-m_2)^2 \;,
\label{k.6} \\[2mm]
&& (m_2 \sp m_3)^2 \leq m_{23}^2 \leq (M-m_1)^2 \;.
\label{k.7} 
\eea

\section{two-meson propagators and functions $\Omega$}
\label{omega} 

Expressions presented here are conventional.
They are displayed for the sake of completeness and rely on the  the results of Ref.\cite{GL85}.
These integrals do not include symmetry factors, which are accounted for in the main text. 
One deals with both $S$ and $P$ waves and the corresponding two-meson propagators are 
associated with the integrals
\bea
&&
\lc I_{ab} ; \,  I_{ab}^{\m \n}  \rc
= \int \frac{d^4  \ell}{(2\p)^4}
\frac{\lc 1; \, \ell^\m \ell^\n \rc} {D_a  D_b },
\label{a.1}\\[2mm]
&& D_a = (\ell \sp p/2)^2 \sm  M_a^2 \;,
\hspace*{6mm} 
D_b = (\ell \sm p/2)^2 \sm  M_b^2 \;,
\label{a.2}
\eea
with $p^2=s$.
Both integrals $I_{ab}$ and  $I_{ab}^{\m\n}   $  are evaluated using dimensional techniques\cite{GL85}.
For $s \geq (M_a \sp M_b)^2 $, the function $I_{ab}$ has the structure
\bea
I_{ab} &\!=\!&  i\;\frac{1}{16 \p^2} \, \lb \Lambda_{ab} + \Pi_{ab} \rb 
\label{a.3}
\eea
where $\Lambda_{ab}  $ is a divergent function of the renormalization scale $\m $
and of the number of dimensions $n\, $, which diverges in the limit $n\rar 4\, $,
whereas $\Pi $ is regular component, given by
\bea
&& \Pi_{ab}(s) =  1 + \frac{m_a^2 \sp m_b^2}{m_a^2 - m_b^2} \, \ln \frac{m_a}{m_b}
-\, \frac{m_a^2 \sm m_b^2}{s} \, \ln\frac{m_a}{m_b}
\nn\\[2mm]
&& - \,\frac{\sqrt{\l}}{s}\;  \ln \lb \frac{s- m_a^2- m_b^2 + \sqrt{\l}} {2\,m_a\,m_b}\rb
+ \,i\,\p\;\frac{\sqrt{\l}}{s} \;,
\label{a.4}\\[4mm]
&& \l = s^2 - 2\;s\;(m_a^2+m_b^2) + (m_a^2-m_b^2)^2 \;.
\label{a.5}
\eea
which, for $a=b\, $, reduces to 
\bea
&& \Pi_{aa}(s) = 2
- \,\frac{\sqrt{\l}}{s}\;  \ln \lb \frac{s- 2\, m_a^2 + \sqrt{\l}} {2\,m_a^2 }\rb
+ \,i\,\p\;\frac{\sqrt{\l}}{s} \;.
\label{a.6}
\eea
The tensor integral is needed for $a=b\, $ only, and one has 
\bea
I_{aa}^{\m\n}  &\!=\!&  i\;\frac{1}{16 \p^2} \, \lc \frac{p^\m \, p^\n}{s} \, 
\lb \Lambda_{aa}^{pp} 
+ \frac{1}{12 } \lb s - 4 \,m_x^2 \,\rb\; \Pi_{aa} \rb 
\right.
\nn\\[4mm]
&\!-\!& \left. g^{\m\n}\, \lb \Lambda_{aa}^g
+ \frac{1}{12 } \lb s - 4 \, m_a^2  \,\rb \; \Pi_{aa} \rb \rc \;,
\label{a.7}
\eea
where $ \Lambda_{aa}^{pp}   $ and $ \Lambda_{aa}^g   $ are divergent quantities.

In the $K$-matrix approximation, one keeps only the imaginary parts of the loop integrals, 
which are contained in the function $\Pi$ and has
\bea
\Pi_{ab} &\!\rar \!& -\, \frac{1}{16\,\p}\, \frac{\sqrt{\l}}{s} \;,
\label{a.8}\\[2mm]
\Pi_{aa}^{\m\n} &\!\rar\!& \frac{1}{192\,\p}\, \lb g^{\m \n} - \frac{p^\m p^\n}{s} \rb \, \frac{\l^{3/2}}{s^2} \;.
\label{a.9}
\eea
In the decay calculation, it is more covenient to use the functions $\Ob$, defined by
\bea
\Pi_{ab} &\!\rar \!& -i\; \Ob_{ab}^S \;,
\label{a.10}\\[2mm]
\Pi_{aa}^{\m\n} &\!\rar\!& \frac{i}{4}\, \lb g^{\m \n} - \frac{p^\m p^\n}{s} \rb \, \Ob_{aa}^P \;.
\label{a.11}
\eea
These results are related with CM momenta by
\bea 
&& \Ob_{ab}^S = -\,\frac{i}{8\p} \; \frac{Q_{ab}}{\sqrt{s}} \;
\theta(s \sm (M_a \sp M_b)^2) \;,
\label{a.12}\\[2mm]
&& \Ob_{aa}^P = -\,\frac{i}{6\p} \; \frac{Q_{aa}^3}{\sqrt{s}} \;
\theta(s \sm4\, M_a^2) \;,
\label{a.13}\\[2mm]
&& Q_{ab} = \frac{1}{2} \, \sqrt{s - 2\,(M_a^2 + M_b^2) + (M_a^2 - M_b^ 2)^2/s} \;,
\label{a.14}
\eea 
where $\theta$ is the Heaviside step function.

\section{partially dressed $\f$ propagator}
\label{phi-propagator}

The bare $\phi$ propagator, $G_{\a\b\g\d}$, is given by  eq.(A.10) of Ref.\cite{EGPR}.
It  is dressed by both $\p \r$ and $\Kb K$ intermediate states and 
the corresponding self-energies are denoted respectively by $\S_{\p\r}$ and $\S_{\Kb K}$.
In this section we consider just contributions of the former kind.
The full propagator is given by 
\bea
i\,\D_{\a\b\g\d} &\!=\!&  i\, \D_{\a\b\g\d}^{(0)} + i\, \D_{\a\b\g\d}^{(1)}  
+ i\, \D_{\a\b\g\d}^{(2)}  + i\, \D_{\a\b\g\d}^{(3)} + \cdots  
\label{dp.1}\\
i\,\D_{\a\b\g\d}^{(0)} &\!=\!&  G_{\a\b\g\d} 
\label{dp.2}\\
i\,\D_{\a\b\g\d}^{(1)} &\!=\!&   G_{\a\b ab}\; \lb -i\, \Sigma^{abcd} \rb\; G_{cd\g\d}
\label{dp.3}\\
i\,\D_{\a\b\g\d}^{(2)} &\!=\!&   G_{\a\b ab}\; \lb -i\, \Sigma^{abef} \rb\; G_{efgh}
\; \lb -i\, \Sigma^{gh cd} \rb\; G_{cd\g\d}
\label{dp.4}
\eea

The $\f \p \r$ interaction is extracted from the lagrangian 
\bea
\cL^{\o_1} &\!=\!& i\,g_1  \; \e^{\m\n\rho\s}\;
\dr^\l \o_{1\, \l\m} \; 
\lb \dr_\n \p^-  \rho_{\rho\s}^+ 
+ \dr_\n \p^+ \rho_{\rho\s}^-
+ \dr_\n \p^0 \rho_{\rho\s}^0 \rb
\label{dp.5}
\eea
where $\o_1= \cos\theta\, \f - \sin\theta \,\o$ is the singlet component.
In the sequence, we write $g_\e = g_1\,\cos\theta$.

\begin{figure}[ht] 
\hspace*{-20mm}
\includegraphics[width=.4\columnwidth,angle=0]{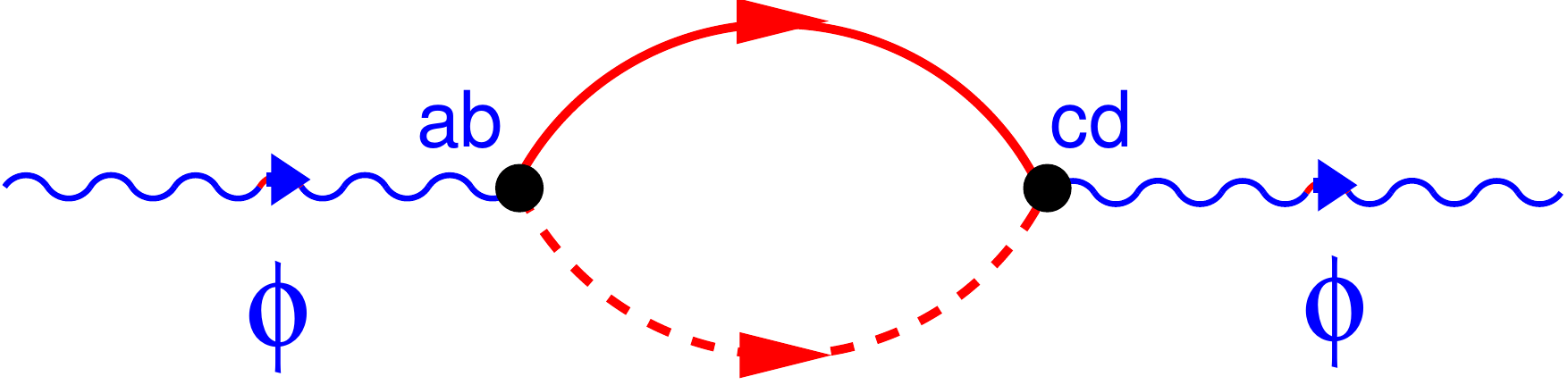}
\caption{Intermediate $\p\r$ contribution to the $\f$ self-energy.}
\label{F1}
\end{figure}

The self energy is given by
\bea
-i\,\Sigma_{\rho\p}^{abcd} &\!=\!& 
\frac{(k^a\,g^{b\m} - g^{a\m}\,k^b)}{2} \; \lb H_{\m\l} \rb \;
\frac{(k^c\,g^{d\l} - g^{c\l}\,k^d)}{2} \;,
\label{dp.6}\\
H_{\m\l}  &\!=\!& 
\lb - 3\, g_\e^2 \; I_{\m\l} \rb \;,
\label{dp.7}\\
I_{\m\l} &\!=\!& \frac{1}{i} \; 
\int \!\! \frac{d^4\ell}{(2\p)^4}\; \frac{p^\m \, p^\l} {p^2 - M_\p^2} \;
\e_{\m\n \x\y} \,G_{\x\y\w\z}(q)\, \e_{\l\xi \w\z} \;,
\label{dp.8}
\eea
with $ p = k/2-\ell, \;\; q =  k/2+\ell$  and $k^2=s\,$.
Using the explicit form of $G_{\x\y\w\z}$ and the definitions
$ D_\p =  p^2 - M_\p^2 \,, \;\; D_\rho = q^2 - \mr2 $, we find
\bea
&& I_{\m\l}  \rar 
\frac{4}{\mr2 } \; 
\int \!\! \frac{d^4\ell}{(2\p)^4}\; \frac{1} {D_\p} \;\frac{1}{D_\r}\;
\label{dp.9}\\
&& \times 
\lc g_{\m\l} \lb -\mr2 \lp M_\p^2 + D_\p  \rp 
+ \frac{1}{4} \,\lp s - M_\p^2 - \mr2 - D_\p - D_\r \rp^2 \rb 
+ \ell_\m \ell_\l \lb k^2 - D_\r \rb \rc \;,
\nn
\eea
where we have used the fact that terms proportional to $k_\m$ and $k_\l$ do not 
contribute to eq.(\ref{dp.6}). 
This integral is highly divergent, but the part regarding the $K \r$ cut is not.
Terms containing factors $D_\p$ and $D_\r$ in the numerator do not contribute 
to the cut function and the relevant integral is 
\bea
I_{\m\l} \!
 \rar \frac{1}{\mr2 } \; 
\int \!\! \frac{d^4\ell}{(2\p)^4}\; \frac{1} {D_\p \; D_\rho }
\lc  \lb  s^2 - 2\, s \, \lp M_\p^2 + \mr2 \rp + \lp M_\p^2 - \mr2 \rp^2 \rb  g_{\m\l}
+ 4\, s \, \ell_\m \ell_\l   \rc \;.
\label{dp.10}
\eea
Using the definition 
\bea
&& I_{\p\rho} = \int \!\! \frac{d^4\ell}{(2\p)^4}\; \frac{1} {D_\p  \; D_\rho } 
\label{dp.11}
\eea
and the result
\bea
\int \!\! \frac{d^4\ell}{(2\p)^4}\; \frac{\ell_\m \,\ell_\l} {D_\p \; D_\rho }\;
&\!=\! & - \lc \frac{1}{12\,k^2} \;
\lb  s^2 - 2\,  s \, (M_\p^2 + \mr2 ) + (M_\p^2 - \mr2 ) ^2 \rb \, I_{\p\rho} \rc g_{\m\l} 
\nn\\
&\! + \!& \mathrm{term} \, \mathrm{proportional} \, \mathrm{to} \, k_\m\,k_\l  \;,
\label{dp.12}
\eea
the relevant component of $I_{\m\l}$ becomes
\bea
I_{\m\l}  &\! \rar \!&  \lc \frac{2}{3\, \mr2 } \; 
\lb  s^2 - 2\, s \, (M_\p^2 + \mr2 ) + (M_\p^2 - \mr2 ) ^2 \rb \; I_{\p\rho} \rc \; g_{\m\l} \;.
\label{dp.13}
\eea
The on-shell contribution to eq.(\ref{dp.11}) is given by 
\bea
I_{\p\rho} 
&\!=\!&  -\; \frac{1}{16\,\p} \frac{\sqrt{\l_{\p \rho}}}{s} \;,
\label{dp.14}
\eea
with 
$ \l_{\p\rho} = \lb  s^2 - 2\, s \, (M_\p^2 + \mr2 ) + (M_\p^2 - \mr2 ) ^2 \rb 
= 4\, s\, Q_{\p\rho}^2  $, where $Q_{\p\rho}$ is the CM three-momentum.
We then have 
\bea
&& H_{\m\l}  =   g_{\m\l} \, \frac{m_\f}{s} \;\G_\f^{\p\r}(s)\;,
\label{dp.15}\\[2mm]
&& m_\f \, \G_\f^{\p\r} (s) = \frac{g_\e^2}{\p\, m_\r^2} \; s^{3/2} \; Q_{\p\rho}^3 \;.
\label{dp.16}
\eea
Numerically, $\G_\f^{\p\r}=0.1532 \times \G_\f=0.1532\times 0.004247\,$GeV\cite{PDG}.
Using this result into eq.(\ref{dp.1}) and ressumming  the series, we get
the partially dressed propagator
\bea
i\,\D_{\a\b\g\d}^{\p\r} &\!=\!&  G_{\a\b\g\d} 
\label{dp.17}\\
&\!+\!&
\lb \frac{i \, m_\f\,\G_\f^{\p\r}(s)/s}{ D_\f^{\p\r}(s) } \rb 
\frac{1}{2}
\lb g_\a^d\,k_\b \,k^c + g_\b^c\,k_\a\,k^d - g_\a^c \, k_\b\,k^d - g_\b^d \,k_\a\,k^c \rb \; 
 G_{cd\g\d}  \;,
\nn
\eea
where the denominator $ D_\f^{\p\r}(s) $ is given by 
\bea
D_\f^{\p\r} &\!=\!&  s-m_\f^2 + i\, m_\f\,\G_\f^{\p\r}(s) \;.
\label{dp.18}
\eea
In the evaluation of amplitudes involving a $\Kb(p_1)\,K(p_2)$ vertex, 
one encounters the product
\bea 
i\, \D_{\a\b\g\d} \, \lp p_1^\g \, p_2^\d - p_2^\g \, p_1^\d \rp   &\!=\!&  
-\, \frac{2\,i}{D_\f^{\p\r}(s)}\; 
\lp p_{1\a} \, p_{2\b} - p_{2\a}\,  p_{1\b} \rp \;.
\label{dp.19}
\eea

\section{SU(3) intermediate states}
\label{states}

In the treatment of intermediate states, it is convenient to work with Cartesian $ SU(3) $ states, 
which are related to charged states by 
\bea
&& | \pi^+ \ra = - | 1 + i \, 2 \ra/\rtw \;,
\hspace{10mm}
| \pi^- \ra = | 1 - i \, 2 \ra/\rtw \;,
\label{su1}\\[2mm]
&& | \pi^0 \ra =  | 3 \ra \;,
\hspace*{30mm}
 | \eta_8 \ra =  | 8 \ra \;,
\label{su2}\\[2mm]
&& | K^+\ra  =  |4 + i \,5 \ra/\rtw \;,
\hspace{10mm}
| K^- \ra = - |4 - i \, 5 \ra/\rtw \;,
\label{su3}\\[2mm]
&& | K^0\ra  = | 6 + i \, 7 \ra/\rtw \;,
\hspace{10mm}
| \bar{K}^0 \ra = |6 - i \, 7 \ra/\rtw \;.
\label{su4}
\eea
We need just two-meson intermediate states $|ab\ra$, with the same quantum numbers as the $ K^-K^+ $ system, 
which are given by
\bea 
|V_3^{\p\p}\ra &\!=\!&  (1/\rtw) \, |1\,2 - 2\,1\ra ,
\label{su5}\\[2mm]
|V_3^{KK} \ra &\!=\!& (1/2) \, | 4\,5 - 5\,4 - 6\,7 + 7\,6 \ra ,
\label{su6}\\[2mm]
| V_8^{KK} \ra &\!=\!&  (1/2) \, | 4\,5  - 5\,4 + 6\,7 - 7\,6 \ra ,
\label{su7}\\[2mm]
| U_3^{\p 8} \ra &\!=\!&  (1/\rtw) \, | 3\, 8 + 8\,3 \ra ,
\label{su8}\\[2mm]
| U_3^{KK} \ra &\!=\!& (1/2) \, | 4\,4 + 5\,5 - 6\,6  - 7\,7 \ra  ,
\label{su9}\\[2mm]
| S^{\p\p} \ra &\!=\!& (1/\rth) \, | 1\,1 + 2\,2 + 3\,3 \ra ,
\label{su10}\\[2mm]
| S^{KK} \ra &\!=\!& (1/2)\, |  4\,4 + 5\,5 + 6\,6 + 7\,7 \ra ,
\label{su11}\\[2mm]
| S^{88} \ra &\!=\!&  |8\, 8\ra  .
\label{su12}
\eea

The state $ |K^- K^+\ra $ includes a conventional phase an reads
\bea
| K^- K^+ \ra &\!=\!& -(1/2) \, | (4 - i\, 5)(4 + i\, 5)\ra
= -(1/2) \, | 4\,4 + 5\, 5\ra  - i\, (1/2) \, | 4\,5-5\,4 \ra 
\label{su9}
\eea
and, therefore,
\bea
\la K^-\,K^+ | &\! =\!& (i/2) \, \la V_3^{KK} + V_8^{KK} | 
- (1/2) \, \la U_3^{KK} + S^{KK} | .
\label{su10} 
\eea


\section{tree decay sub-amplitudes}
\label{treedec}

In the evaluation of intermediate state contributions shown in diagrams of Fig.\ref{modT},
we need tree level contribution for the process $D\rar  a\, b \, K^+ $,
denoted by $ T_{(0)}^{(J,I)} $, for spin $ J $ and isospin $ I $.
In the results displayed below, the first terms correspond to resonances in  
diagrams (3A+3B), whereas those within square brackets, labeled by $c$, represent 
contact interactions in the top vertices of diagrams 2A and 2B.
Using the constant  $ C $ defined in eq.(\ref{dec1a}), we have
\bea
&& [J, I =1,1]  \rar \la V_3^{a b}\, K^+ |\, T_{(0)}^{(1,1)} \,| \, D \ra 
= \frac{i}{2}\, (m_{13}^2-m_{23}^2) \; \G_{(0)\,a\,b}^{(1,1)} \,,
\label{tsa1}\\[4mm]
&& \G_{(0)\,\p\p}^{(1,1)}  =  C \lc 
\lb \frac{\rtw\, G_V^2}{ F^2} \rb \, \frac{m_{12}^2}{m_{12}^2 - m_\rho^2} 
+ \lb -\,  \frac{1}{\rtw}\rb_c \rc ,
\label{tsa2} \\[4mm]
&& \G_{(0)\, KK}^{(1,1)}  = C \lc 
\lb \frac{G_V^2}{F^2} \rb \, \frac{m_{12}^2}{m_{12}^2 - m_\rho^2} 
+ \lb - \,  \frac{1}{2} \rb_c \rc
\label{tsa3}\\[4mm]
&& [J, I =1,0] \rar \la V_8^{KK} \, K^+|\, T_{(0)}^{(1,0)} \,| \, D \ra 
= \frac{i}{2}\, (m_{13}^2 - m_{23}^2) \; \G_{(0)\,KK}^{(1,0)}  \,,
\label{tsa4}\\[4mm]
&& \G_{(0)\,KK}^{(1,0)} = C \lc 
\lb \frac{3\,G_V^2}{F^2}\, \sin^2\!\theta \rb \, \frac{m_{12}^2}{D_\f^{\p\rho}(m_{12}^2)} 
+ \lb -\,  \frac{3}{2}\rb_c \rc ,
\label{tsa5}
\eea
Here, the function $ D_\f^{\p\rho} $ is a partially dressed $\phi$ propagator, 
discussed in App.\ref{phi-propagator}, eq.({\ref{dp.18}),
associated with the partial width of the decay $\phi \rar (\rho \p + \p\p\p)$. 

\bea
&& [J, I =0,1] \rar \la U_3^{ab} \, K^+|\, T_{(0)}^{(0,1)} \,| \, D \ra = \G_{(0)\, a\,b}^{(0,1)} \,,
\label{tsa6}\\[4mm]
&& \G_{(0)\,\p 8}^{(0,1)}  = C \lc 
\lb \frac{2\,\rtw}{\rth\,F^2} \rb  \;
\frac{ \lb - c_d \, P\cd p_3 + c_m \, M_D^2 \rb}{m_{12}^2 - m_{a_0}^2} 
\lb c_d \, \lp m_{12}^2 - M_\p^2 - M_8^2 \rp  
 + 2\, c_m \, M_\p^2 \rb 
 \right.
 \nn   \\[2mm]
 && \left.
+ \lb -\, \frac{\rth}{\rtw} \, \lb \, M_D^2/3 - P\cd p_3 \rb \rb_c \rc ,
 \label{tsa7} \\[4mm]
&& \G_{(0)\,KK}^{(0,1)}  = C \lc 
\lb \frac{2}{F^2} \rb \;
\frac{ \lb - c_d \, P\cd p_3 + c_m \, M_D^2 \rb}{m_{12}^2 - m_{a_0}^2}
\lb c_d \, \lp m_{12}^2 - 2 M_K^2 \rp  
 + 2\, c_m \, M_K^2 \rb 
 \right.
 \nn\\[2mm]
 &&  \left.
+\lb -\, \frac{1}{2} \lb M_D^2 -  P \cd p_3 \rb \rb_c\rc ,
\label{tsa8} \\[4mm] 
&& [J, I =0,0] \rar \la S^{ab} \, K^+|\, T_{(0)}^{(0,0)} \,| \, D \ra = \G_{(0)\,a\,b}^{(0,0)} \,,
\label{tsa9}\\[4mm]
&& \G_{(0)\, \p\p}^{(0, 0)}  = C \lc 
\lb \frac{8 \rth}{F^2} \rb \; 
\frac{ \lb -\ct_d \, P\cd p_3 + \ct_m \, M_D^2 \rb}{m_{12}^2 - m_{S1}^2}\;
\lb \ct_d\, \lp m_{12}^2 - 2 M_\p^2 \rp  
 + 2\, \ct_m \, M_\p^2 \rb 
 \right.
\nn\\[4mm]
&&  \left. 
- \;  \lb \frac{2}{\rth \, F^2} \rb  \;
\frac{ \lb - c_d \, P\cd p_3 + c_m \, M_D^2 \rb}{m_{12}^2 - m_{So}^2} \;
\lb c_d \, \lp m_{12}^2 - 2 M_\p^2 \rp  
 + 2\, c_m \, M_\p^2 \rb 
 \right.
 \nn\\[2mm]
 && \left.
+\lb -\, \frac{\rth}{2} \, \lb M_D^2 - P\cd p_3 \rb \rb_c  \rc ,
\label{tsa10} \\[4mm]
&& \G_{(0)\,KK}^{(0,0)}  =  C \lc 
\lb \frac{16}{F^2} \rb \; 
\frac{ \lb - \ct_d \, P\cd p_3 + \ct_m \, M_D^2 \rb}{m_{12}^2 - m_{S1}^2}\;
\lb \ct_d \, \lp m_{12}^2 - 2 M_K^2 \rp  
 + 2\, \ct_m \, M_K^2 \rb 
 \right.
\nn\\[4mm]
&& \left.
+ \; \lb \frac{2}{3\, F^2} \rb  \;
\frac{ \lb -c_d \, P\cd p_3 + c_m \, M_D^2 \rb}{m_{12}^2 - m_{So}^2}
\lb c_d \, \lp m_{12}^2 - 2 M_K^2 \rp  
 + 2\, c_m \, M_K^2 \rb 
 \right.
 \nn\\[2mm]
 && \left.
+ \lb -\, \frac{3}{2} \, \lb M_D^2 - P\cd p_3 \rb \rb_c \rc ,
\label{tsa11}  \\[4mm] 
%
%
&& \G_{(0)\, 88}^{(0,0)}  = C \lc 
 \lb \frac{8}{F^2} \rb \;
\frac{ \lb -\ct_d \, P\cd p_3 + \ct_m \, M_D^2 \rb}{m_{12}^2 - m_{S1}^2}\;
\lb \ct_d \, \lp m_{12}^2 - 2 M_8^2 \rp  
 + 2\, \ct_m \, M_8^2  \rb 
 \right.
 \nn\\[4mm]
 && \left. 
 + \; \lb \frac{2}{3\, F^2} \rb  \;
\frac{ \lb -c_d \, P\cd p_3 + c_m \, M_D^2 \rb}{m_{12}^2 - m_{So}^2}
\lb c_d \, \lp m_{12}^2 - 2 M_8^2 \rp  
 + c_m \, \lp - 10 M_\p^2 + 16M_K^2 \rp/3 \rb 
 \right.
 \nn\\[2mm]
 && \left.
 + \lb - \, \frac{1}{2} \, \lb 5\,M_D^2/3  - 3\, P \cd p_3 \rb \rb_c \rc .
\label{tsa12} 
\eea
with
\bea 
P\cd p_3 = \frac{1}{2} \lb M_D^2 + M_K^2 - m_{12}^2\rb \,.
\label{tsa13}
\eea

\section{scattering kernels }
\label{kernel}

The intermediate scattering amplitudes depend on interaction kernels in the four channels considered, 
associated with $J,I=1,0 $.
The kernel matrix elements for the reaction  $ c\,d \rar a\, b $ are written as 
$ \la cd \,| \, \cK^{J,I} \, |\, ab \ra $, in terms of the states defined in App.\ref{states}, 
and  displayed below.
All kernels are written as sums of NLO resonance contributions and chiral polynomials,
involving both LO and NLO terms.
The NLO polynomials are derived by assuming that the LECs are saturared by intermedate
vector and scalar resonances, with masses $ M_V $ and $ M_S $, respectively.
The kernel matrix elements read
\bea 
&& [J, I =1,1] \rar \la V_3^{ab}\,| \,\cK^{(1,1)}\,|\,V_3^{cd}\ra 
=  (t\sm u)\;  \cK_{(ab|cd)}^{(1,1)}  
\label{k1}\\[4mm]
&& \cK_{(\p \p|\p\p)}^{(1,1)} =   - 2\, \lb\frac{G_V^2}{F^4} \rb \;\frac{s}{s-m_\rho^2} 
+ \lb \frac{1}{F^2} \rb_c
\label{k2}\\[4mm]
&& \cK_{(\p \p|KK)}^{(1,1)} = - \rtw \, \lb\frac{G_V^2}{F^4} \rb \;\frac{s}{s-m_\rho^2} 
+  \lb \frac{\rtw}{2\,F^2} \rb_c 
\label{k3}\\[4mm]
&& \cK_{(KK|KK)}^{(1,1)} = - \lb\frac{G_V^2}{F^4} \rb \;\frac{s}{s-m_\rho^2} 
+    \lb \frac{1}{2\,F^2} \rb_c 
\label{k4} 
\eea

\bea 
&& [J,I=1,0] \rar  \la V_8^{ab}\,| \,\cK^{(1,0)}\,|\,V_8^{cd}\ra 
= (t \sm u) \; \cK_{(ab|cd)}^{(1,0)}   
\label{k5}\\[4mm]
&& \cK_{(KK|KK)}^{(1,0)} = - 3\, \lb\frac{G_V^2 \; \sin^2\!\theta}{F^4} \rb \;\frac{s}{D_\phi^{\p\rho}} 
+  \lb \frac{3}{2\,F^2} \rb_c
\label{k6}
\eea
The function $ D_\f^{\p\rho} $ is this expression represents a partially dressed $\phi$ propagator, 
discussed in App.\ref{phi-propagator}, eq.({\ref{dp.18}),
and accounts for the partial width of the decay $\phi \rar (\rho \p + \p\p\p)$. 

\bea 
&& [J,I=0,1] \rar \la U_3^{ab} \,| \,\cK^{(0,1)}\,|\,U_3^{cd}\ra  = \cK_{(ab|cd)}^{(0,1)}
\label{k7}\\[4mm]
&& \cK_{(\p 8|\p 8)}^{(0,1)} =  -\, \frac{1}{s-m_{a_0}^2}\,\lb\frac{4}{3\,F^4} \rb \;
\lb c_d \, (s\sm M_\p^2 \sm M_8^2)  + c_m \, 2M_\p^2 \rb^2 
+  \lb \frac{2 M_\p^2}{3 F^2} \rb_c 
\label{k8}
\\[4mm]
&& \cK_{(\p 8|KK)}^{(0,1)} =  -\, \frac{1}{s-m_{a_0}^2}\,\lb\frac{2\rtw}{\rth\,F^4} \rb 
\lb c_d \, (s\sm M_\p^2 \sm M_8^2)  + c_m \, 2M_\p^2 \rb \,
\lb c_d \, s - (c_d \sm c_m) \, 2M_K^2 \rb 
\nn\\[2mm]
&& + \lb \frac{(3s -4M_K^2)}{\rts\, F^2} \rb_c 
\label{k9}\\[4mm]
&& \cK_{(KK|KK)}^{(0,1)}  =  -\frac{1}{s-m_{a_0}^2}\,\lb\frac{2}{F^4} \rb \;
\lb c_d \, s - (c_d \sm c_m) \, 2M_K^2 \rb^2 
+ \lb \frac{s}{2F^2} \rb_c 
\label{k10}
\eea

\bea 
&& [J,I=0,0] \rar \la S^{ab} \,| \,\cK^{(0,0)}\,|\,S^{cd}\ra 
= \cK_{(ab|cd)}^{(0,0)} 
\label{k11}\\[4mm]
&& \cK_{(\p \p|\p \p)}^{(0,0)}  = -\, \frac{1}{s-m_{S1}^2}\,\lb\frac{12}{F^4} \rb \;
\lb \ct_d \, s - (\ct_d \sm \ct_m) \, 2M_\p^2\rb^2 
\nn\\[2mm]
&& -\, \frac{1}{s-m_{So}^2}\,\lb\frac{2}{F^4} \rb \;
\lb c_d \, s - (c_d \sm c_m) \, 2M_\p^2 \rb^2 
+ \lb \frac{2 s - M_\p^2}{ F^2} \rb_c
\label{k12}\\[4mm]
%
%
%
%
&& \cK_{(\p \p|KK)}^{(0,0)}  = -\,\frac{1}{s-m_{S1}^2}\,\lb\frac{8\rth}{F^4} \rb \;
\lb \ct_d \, s - (\ct_d \sm \ct_m) \, 2M_\p^2\rb \,
\lb \ct_d \, s - (\ct_d \sm \ct_m) \, 2M_K^2 \rb 
\nn\\[2mm]
&& +   \frac{1}{s-m_{So}^2}\,\lb\frac{2}{\rth\,F^4} \rb \;
\lb c_d \, s - (c_d \sm c_m) \, 2M_\p^2 \rb 
\lb c_d \, s - (c_d \sm c_m) \, 2M_K^2 \rb  
+  \lb \frac{\rth\, s}{2 F^2} \rb_c
\label{k13}\\[4mm]
%
%
%
%
&& \cK_{(\p \p|88)}^{(0,0)}  = -\, \frac{1}{s-m_{S1}^2}\,\lb\frac{4\rth}{F^4} \rb \;
\lb \ct_d \, s - (\ct_d \sm \ct_m) \, 2M_\p^2 \rb \,
\lb \ct_d \, s - (\ct_d \sm \ct_m) \, 2M_8^2\rb 
\nn\\[2mm]
&& +  \frac{1}{s-m_{So}^2}\,\lb\frac{2}{\rth\,F^4} \rb \;
\lb c_d \, s - (c_d \sm c_m) \, 2M_\p^2 \rb 
\lb c_d \, (s\sm 2 M_8^2)  + c_m \, (16M_K^2 \sm 10 M_\p^2)/3\rb 
\nn\\[2mm]
&& + \,  \lb \frac{\rth\, M_\p^2}{3 F^2} \rb_c
\label{k14}\\[4mm]
%
%
%
&& \cK_{(KK|KK)}^{(0,0)}  = -\, \frac{1}{s-m_{S1}^2}\,\lb\frac{16}{F^4} \rb \;
\lb \ct_d \, s - (\ct_d \sm \ct_m) \, 2M_K^2\rb^2 
\nn\\[2mm]
&& -   \frac{1}{s-m_{So}^2}\,\lb\frac{2}{3\,F^4} \rb \;
\lb c_d \, s - (c_d \sm c_m) \, 2 M_K^2 \rb^2 
+  \lb  \frac{3 s}{2 F^2}  \rb_c
\label{k15}\\[4mm]
%
%
%
%
&& \cK_{(KK|88)}^{(0,0)}  =  -\, \frac{1}{s-m_{S1}^2}\,\lb\frac{8}{F^4} \rb \;
\lb \ct_d \, s - (\ct_d \sm \ct_m) \, 2M_K^2 \rb \,
\lb \ct_d \, s - (\ct_d \sm \ct_m) \, 2M_8^2 \rb 
\nn\\[2mm]
&& -  \, \frac{1}{s-m_{So}^2}\,\lb\frac{2}{3\,F^4} \rb \;
\lb c_d \, s - (c_d \sm c_m) \, 2 M_K^2 \rb 
\lb c_d \, (s\sm 2 M_8^2)  + c_m \, (16M_K^2 \sm 10 M_\p^2)/3\rb 
\nn \\[2mm]
&& +  \,  \lb  \frac{9 s - 8M_K^2}{6 F^2} \rb_c
\label{k16}\\[4mm]
%
%
%
%
&& \cK_{(88|88)}^{(0,0)}  =  -\, \frac{1}{s-m_{S1}^2}\,\lb\frac{4}{F^4} \rb \;
\lb \ct_d \, s - (\ct_d \sm \ct_m) \, 2M_8^2\rb^2 
\nn\\[2mm]
&& -  \, \frac{1}{s-m_{So}^2}\,\lb\frac{2}{3\, F^4} \rb \;
\lb c_d \, (s\sm 2 M_8^2)  + c_m \, (16M_K^2 \sm 10 M_\p^2)/3\rb^2 
\nn\\[2mm]
&& 
+ \,  \lb \frac{-7M_\p^2 +16 M_K^2}{9 F^2} \rb_c 
\label{k17}
\eea

\section{channel dependent decay amplitudes - full results}
\label{decayT} 
 
The tree level decay amplitudes for channel with spin $ J $ and isospin $ I $, 
given in App.\ref{treedec},  are written as  
\bea
\la X^{a b}\, K^+ |\, T_{(0)}^{(J,I)} \,| \, D \ra 
&\!=\!&  \frac{i}{2}\, (m_{13}^2 - m_{23}^2)  \; \G_{(0)\,a\,b}^{(1,I)}  
\;\;\; \rar (X=V_3, V_8)
\nn\\[2mm]
&\!=\!&  \G_{(0)\,a\,b}^{(0,I)} 
\;\;\; \rar  (X=U_3, S) 
\label{z1}
\eea
The full amplitudes are obtained by including all possible final state interactions, 
as indicated in Figs.\ref{modT} and \ref{modA}.
The terms involving a single meson-meson interaction read
\bea
\la X^{a b}\, K^+ |\, T_{(1)}^{(J,I)} \,| \, D \ra 
&\!=\!&  \frac{i}{2}\, (m_{13}^2 -m_{23}^2) \; \G_{(1)\,a\,b}^{(1,I)}  
\;\;\; \rar (X=V_3, V_8)
\nn\\[2mm]
&\!=\!&  \G_{(1)\,a\,b}^{(0,I)} 
\;\;\; \rar  (X=U_3, S) 
\label{z2}
\eea
with 
\bea
&& \G_{(1)\,ab }^{(J,I)} = \sum_{cd} \; \,  M_{ab|cd}^{(J,I)} \; \G_{(0)\,cd}^{(J,I)} \,,
\label{z3}\\[2mm]
&& M_{ab|cd}^{(J,I)} = - \cK_{ab|cd}^{(J,I)} \, \lb S_F \, \Ob_{cd}^J \rb  \;.
\label{z4}
\eea
where $ \cK_{ab|cd}^{(J,I)}  $  are the scattering kernels 
displayed in App.\ref{kernel}, 
$\Ob_{cd}^J$  are the two-meson propagators given in App.\ref{omega}, and the 
symmetry factor $ S_F = 1 \rar c \neq d $  and  $ S_F = 1/2  \rar c = d $.
The terms $\G_{(2)\,a\,b}^{(J,I)}$,  containing two meson-meson interactions 
are constructed in a similar way from $\G_{(1)\,a\,b}^{(J,I)}$, and so on.

The inclusion of all possible meson-meson interactions leads to the infinite geometric series
\bea
&& \G_{ab}^{(J,I)} = \s_{ab|cd}^{(J,I)} \;  \G_{(0)\, cd}^{(J,I)}  \;,
\label{z5}\\
&& \s_{ab|cd}^{(J,I)} = \lc 1 +  M^{(J,I)} + [M^{(J,I)}]^2 + \cdots \rc_{ab|cd} \;,
\label{z6}
\eea
where $ \s^{(J,I)} $ is its  sum, given by
\bea
\s^{(J,I)} = \lb 1 - M^{(J,I)} \rb^{-1} \;.
\label{z7}
\eea
Thus, decay amplitude reads formally
\bea
&& \G^{(J,I)} =  \lb 1 - M^{(J,I)} \rb^{-1} \, \G_{(0)}^{(J,I)} \;.
\label{z8}
\eea
and encompasses a coupled channel structure, which depends on the 
spin-isospin considered.

In order to display the meaning of the indices  used in this structure, 
we label informally each $(J,I)$ channel by its most prominent resonance 
and recall that
\\
$\rho$-channel: $ \G_{(0)\, 11}^{(1,1)} \se \G_{(0)\,\p\p}^{(1,1)}\,,
\G_{(0)\, 22}^{(1,1)} \se \G_{(0)\,KK}^{(1,1)} \,$;
\\[1mm]
$\phi$-channel:
$\G_{(0)}^{(1,0)} \se \G_{(0)\,KK}^{(1,0)}  \,$;
\\[1mm]
$a_0$-cannel:
$\G_{(0)\,11}^{(0,1)} \se \G_{(0)\,\p 8}^{(0,1)}  \,,
\G_{(0)\,22}^{(0,1)} \se \G_{(0)\,KK}^{(0,1)}  \,$;
\\[1mm]
$f_0$-channel:
$\G_{(0)\,11}^{(0,0)} \se \G_{(0)\,\p\p}^{(0,0)} \,,
\G_{(0)\, 22}^{(0,0)} \se \G_{(0)\,KK}^{(0,0)}  \,,
\G_{(0)\,33}^{(0,0)} \se  \G_{(0)\,88}^{(0,0)}  \,$.
\\[1mm]
The meanings of the indices used in the matrices $M^{(J,I)}$, eq.(\ref{z4}), 
are similar.

In this work, we need at most three coupled channels, which corresponds to 
\bea 
&& \hspace*{-5mm}
\s = \frac{1}{\det [1-M]} 
\nn \\[2mm]
&&  \hspace*{-5mm}
\times \lb \matrix{[1\sm M_{22}][1\sm M_{33}] \sm M_{23} M_{32}
& M_{12}[1\sm M_{33}] \sp M_{13} M_{32}
& M_{13}[1\sm M_{22}] \sp M_{12} M_{23}
\cr  
M_{21}[1\sm M_{33}] \sp M_{23} M_{31}
& [1\sm M_{11}][1\sm M_{33}] \sm M_{13} M_{31} 
& M_{23}[1\sm M_{11}] \sp M_{13} M_{21}
\cr
M_{31}[1\sm M_{22}] \sp M_{21} M_{32}
&  M_{32}[1\sm M_{11}] \sp M_{12} M_{31}
& [1\sm M_{11}][1\sm M_{22}] \sm M_{12} M_{21} 
}\rb 
\nn\\[4mm]
&&  \hspace*{-5mm}
\det (1-M) = [1\sm M_{11}][1\sm M_{22}][1 \sm M_{33}]
-  [1 \sm M_{11}]  M_{23} M_{32} 
- [1 \sm M_{22}] M_{13} M_{31} 
\nn\\
&&  
- [1 \sm M_{33}] M_{12} M_{21} 
-  M_{12}M_{23}M_{31} 
- M_{21}M_{32}M_{13}
\label{z9}
\eea

In the $K$-matrix approximation, the matrix elements $M$ are purely imaginary, 
owing to the presence of the two-meson propagator.
The explicit functions  to be used in the calculation are displayed below.
\bea
&& M_{11}^{(1,1)} =  - \cK_{\p\p|\p\p}^{(1, 1)} \, [ \Ob_{\p\p}^P /2 ]  \;,
\hspace{10mm}
M_{12}^{(1,1)} =  -  \cK_{\p\p|KK}^{(1, 1)}  \, [\Ob_{KK}^P/2] \;,
\nn \\[2mm]
&& M_{21}^{(1,1)}  = - \cK_{\p\p|KK}^{(1, 1)} \, [\Ob_{\p\p}^P/2] \;,
\hspace{10mm}
M_{22}^{(1,1)} = -  \cK_{KK|KK}^{(1, 1)} \, [\Ob_{KK}^P /2] \;.
\label{z10}
\eea
\bea
&& M^{(1,0)} = - \cK_{KK|KK}^{(1, 0)} \, [\Ob_{KK}^P/2]\;. 
\label{z11}
\eea
\bea
&& M_{11}^{(0,1)} =  - \cK_{\p 8|\p 8}^{(0, 1)} \, [\Ob_{\p 8}^S/2] \;,
\hspace{10mm} 
M_{12}^{(0,1)} =  - \cK_{\p 8 |KK}^{(0, 1)} \, [\Ob_{KK}^S/2] \;,
\nn \\[2mm]
&& M_{21}^{(0,1)}  = - \cK_{\p 8|KK}^{(0, 1)} \, [\Ob_{\p8}^S/2] \;,  
\hspace{10mm}
M_{22}^{(0,1)} = - \cK_{KK|KK}^{(0, 1)} \, [\Ob_{KK}^S/2 ]\;. 
\label{z12}
\eea
\bea
&& M_{11}^{(0,0)} =  - \cK_{\p\p|\p\p}^{(0,0)} \, [\Ob_{\p\p}^S/2] \;, 
\hspace{10mm}
M_{12}^{(0,0)} =  - \cK_{\p\p|KK}^{(0,0)} \, [\Ob_{KK}^S/2] \;,
\nn\\[2mm]
&& M_{13}^{(0,0)} =  - \cK_{\p\p|88}^{(0,0)} \, [\Ob_{88}^S/2] \;,
\hspace{10mm}
M_{21}^{(0,0)} =  - \cK_{\p\p|KK}^{(0,0)} \, [\Ob_{\p\p}^S/2] \;,
\nn\\[2mm]
&& M_{22}^{(0,0)} =  - \cK_{KK|KK}^{(0,0)} \, [\Ob_{KK}^S/2] \;,
\hspace{10mm}
M_{23} ^{(0,0)}=  - \cK_{KK|88}^{(0,0)} \, [\Ob_{88}^S/2] \;,
\nn \\[2mm]
&& M_{31}^{(0,0)} =  - \cK_{\p\p|88}^{(0,0)} \, [\Ob_{\p\p}^S/2] \;,
\hspace{10mm}
M_{32}^{(0,0)} =  - \cK_{KK|88}^{(0,0)} \, [\Ob_{KK}^S/2] \;,
\nn \\[2mm]
&& M_{33}^{(0,0)} =  - \cK_{88|88}^{(0,0)} \, [\Ob_{88}^S/2] \;.
\label{z13}
\eea
The factor $1/2$ accounts for the symmetry of intermediate states.
It is also present in the functions $M_{11}^{(0,1)}$ and $M_{21}^{(0,1)}$
because one is using the symmetrized $\pi 8$ intermediate state 
given by eq.(\ref{su8}).

In the evaluation of the channel dependent decay amplitudes,  
one subtracts contributions already included in the non-resonant term, 
so as to avoid double counting.
These terms are denoted by $ \G_{c|KK}^{(J,I)} $ and correspond to the contributions 
denoted by $ [\cdots]_c$ in App.\ref{treedec}.
Explicit expressions for the vector channel read
\bea
&& T^{(1,1)}  = -\, \frac{1}{4}\, \lb \Gb_{KK}^{(1,1)}  - \G_{c|KK}^{(1,1)} \rb  \, (m_{13}^2 \sm m_{23}^2)\,,
\label{z14} \\[2mm]
&& \Gb_{KK}^{(1,1)} = \frac{1}{D_\rho (m_{12}^2)}
\lb M_{21}^{(1,1)}  \, \G_{(0)\, \p\p}^{(1,1)} + \lp 1 \sm M_{11}^{(1,1)} \rp \, \G_{(0)\, KK}^{(1,1)}  \rb \,,
\label{z15} \\[4mm] 
&& D_\rho = 
\lb \lp 1\sm M_{11}^{(1,1)} \rp \,  \lp 1\sm M_{22}^{(1,1)} \rp - M_{12}^{(1,1)}  M_{21}^{(1,1)}  \rb \,.
\label{z16}  
\eea
\bea
&& T^{(1,0)}  = -\,  \frac{1}{4}\,
\lb \Gb_{KK}^{(1,0)}  - \G_{c|KK}^{(1,0)}  \rb  \, (m_{13}^2 \sm m_{23}^2) \,,
\label{z17} \\[2mm]
&& \Gb_{KK}^{(1,0)} 
= \frac{1}{D_\phi (m_{12}^2)}\; \G_{(0)\, KK}^{(1,0)} \,,
\label{z18} \\[4mm]
&& D_\phi =  \lc 1\sm M^{(1,0)} \rc \,.
\label{z19} 
\eea
The function $D_\f^{\p\rho}$ in these results is given by eq.(\ref{dp.18})
and corresponds to the part of the $\phi$ propagator involving $\pi \rho$ 
intermediate states. 

The scalar sector yields 
\bea
&& T^{(0,1)}  =  -\, \frac{1}{2}\, \lb \Gb_{KK}^{(0,1)}  - \G_{c|KK}^{(0,1)} \rb  \;,
\label{z20} \\[2mm]
&& \Gb_{KK}^{(0,1)} = \frac{1}{D_{a_0} (m_{12}^2)}
\lb M_{21}^{(0,1)}  \, \G_{(0)\, \p 8}^{(0,1)} + \lp 1 \sm M_{11}^{(0,1)} \rp \, \G_{(0)\, KK}^{(0,1)}  \rb \,,
\label{z21} \\[4mm] 
&& D_{a_0} = 
\lb \lp 1\sm M_{11}^{(0,1)} \rp \,  \lp 1\sm M_{22}^{(0,1)} \rp - M_{12}^{(0,1)}  M_{21}^{(0,1)}  \rb \,,
\label{z22}  
\eea

\bea
&& T^{(0,0)}  =  -\, \frac{1}{2}\, \lb \Gb_{KK}^{(0,0)}  - \G_{c|KK}^{(0,0)} \rb  \,,
\label{z23}\\[2mm]
&& \Gb_{KK}^{(0,0)} = \frac{1}{D_S(m_{12}^2) }
\lc \lb M_{21}^{(0,0)} \lp 1\sm M_{33}^{(0,0)}\rp \sp M_{23}^{(0,0)} M_{31}^{(0,0)}\rb \, \G_{(0)\,\p\p}^{(0,0)} 
\right.
\nn \\[2mm]
&& \left. + \lb \lp 1\sm M_{11}^{(0,0)}\rp \lp 1\sm M_{33}^{(0,0)}\rp  
\sm M_{13}^{(0,0)} M_{31}^{(0,0)}\rb  \, \G_{(0)\,KK}^{(0,0)} 
\right.
\nn\\[2mm]
&&  \left. +\, 
\lb  M_{23}^{(0,0)} \lp 1\sm M_{11}^{(0,0)} \rp  \sp M_{13}^{(0,0)} M_{21}^{(0,0)} \rb  \, \G_{(0)\,88}^{(0,0)} \rc \;,
\label{z24}\\[2mm]
&& D_S = \det \lb 1-M^{(0,0)}\rb \;.
\label{z25}
\eea

\section{channel dependent scattering amplitudes - full results}
\label{scatt} 

The scattering amplitudes for channels with spin $ J $ and isospin $ I $ are given by  
\bea
\la X^{ab} |\, A \,| X^{cd} \ra &\!=\!& (t-u)\, A_{ab|cd}^{(1,I)}
\;\;\; \rar (X=V_3, V_8) \;,
\nn \\[2mm]
\la X^{ab} |\, A \,| X^{cd} \ra &\!=\!& A_{ab|cd}^{(0,I)}
\;\;\; \rar  (X=U_3, S) \;,
\label{w1}
\eea
whereas the tree approximation reads
\bea
\la X^{ab} |\, A_{(0)} \,| X^{cd} \ra &\!=\!& (t-u)\, \cK_{ab|cd}^{(1,I)}
\;\;\; \rar (X=V_3, V_8) \;,
\nn \\[2mm]
\la X^{ab} |\, A_{(0)} \,| X^{cd} \ra &\!=\!& \cK_{ab|cd}^{(0,I)}
\;\;\; \rar  (X=U_3, S) \;,
\label{w2}
\eea
with the $\cK$ given in App.\ref{kernel}.
The full amplitudes are obtained by including all loop contributions, as indicated in Fig.\ref{modA}.
The terms involving a single loop read
\bea 
&& A_{(1)\, ab|cd}^{(J,I)} 
=\sum_{ef} \;  M_{ab|ef}^{(J,I)} \; A_{(0)\,ef|cd}^{(J,I)}
\label{w3}\\[2mm]
&& M_{ab|ef}^{(J,I)} = - \cK_{ab|ef}^{(J,I)} \, \lb S_F \, \Ob_{ef}^J \rb  \;.
\label{w4}
\eea
where the $\Ob_{ef}^J$  are the two-meson propagators given in App.\ref{omega}, with the 
symmetry factor $ S_F = 1 \rar e \neq f $  and  $ S_F = 1/2  \rar e = f $.
The inclusion of all possible intermediate loops gives rise to the infinite geometric series
\bea
&& A_{ab|cd}^{(J,I)} = \s_{ab|ef}^{(J,I)} \;  A_{(0)\, ef|cd}^{(J,I)}  \;,
\label{w5}\\
&& \s_{ab|ef}^{(J,I)} = \lc 1 +  M^{(J,I)} + [M^{(J,I)}]^2 + \cdots \rc_{ab|ef} \;,
\label{w6}
\eea
which is very similar to that discussed in eq.(\ref{z5}).
In particular, the function $\s_{ab|ef}^{(J,I)} $ is the same as eq.(\ref{z6})  
and therefore we may rely on all the developments made in App.\ref{decayT}.
Explicit expressions for the vector scattering amplitudes read
\bea
A_{KK|KK}^{(1,1)} &\!=\!& \frac{1}{D_\rho (m_{12}^2)}
\lb M_{21}^{(1,1)} \, \cK_{\p\p|KK}^{(1,1)} + \lp 1 \sm M_{11}^{(1,1)}\rp \, \cK_{KK|KK}^{(1,1)}  \rb \;,
\label{w7}\\[4mm] 
D_\rho &\!=\!& 
\lb \lp 1\sm M_{11}^{(1,1)}\rp  \,  \lp 1\sm M_{22}^{(1,1)}\rp  - M_{12}^{(1,1)}\, M_{21}^{(1,1)} \rb \;,
\label{w8}\\[4mm]
A_{KK|KK}^{(1,0)} &\!=\!&  \frac{1}{D_\phi (m_{12}^2)}\;  \cK_{KK|KK}^{(1,0)} \;,
\label{w9}\\[4mm]
D_\phi &\!=\!&  \lc 1\sm M^{(1,0)} \rc \,,
\label{w10} 
\eea
where the function $D_\f^{\p\rho}$ is given by eq.(\ref{dp.18}).

The scalar sector yields 
\bea
A_{KK|KK}^{(0,1)} &\!=\!& \frac{ 1}{D_{a_0} (m_{12}^2)}
\lb M_{21}^{(0,1)} \,  \cK_{\p 8|KK}^{(0,1)} + \lp 1 \sm M_{11}^{(0,1)} \rp \, \cK_{ KK|KK}^{(0,1)}  \rb 
\label{w11}\\[4mm] 
D_{a_0} &\!=\!&  \, 
\lb \lp 1\sm M_{11}^{(0,1)} \rp \,  \lp 1\sm M_{22}^{(0,1)} \rp - M_{12}^{(0,1)}  M_{21}^{(0,1)}  \rb \,,
\label{w12} \\[4mm]
A_{KK|KK}^{(0,0)} &\!=\!& \frac{1}{D_S (m_{12}^2)}
\lc \lb M_{21}^{(0,0)} \lp 1\sm M_{33}^{(0,0)} \rp \sp M_{23}^{(0,0)} M_{31}^{(0,0)}\rb  \, \cK_{\p\p|KK}^{(0,0)}
\right.
\nn\\[2mm]
&\!+\!& \left.  
\lb \lp 1\sm M_{11}^{(0,0)} \rp \lp1\sm M_{33}^{(0,0)} \rp \sm M_{13}^{(0,0)} M_{31}^{(0,0)}\rb \, \cK_{KK|KK}^{(0,0)} 
\right.
\nn\\[2mm]
&\!+\!& \left.  
\lb M_{23}^{(0,0)} \lp 1\sm M_{11}^{(0,0)} \rp \sp M_{13}^{(0,0)} M_{21}^{(0,0)}\rb  \, \cK_{88|KK}^{(0,0)} \rc \;,
\label{w13}\\[2mm]
D_S &\!=\!& \det \lp 1-M^{(0,0)} \rp \;,
\label{w14}
\eea
with $ \det \lp 1-M^{(0,0)} \rp $ given by eq.(\ref{z9}).

\section{phase shifts}

The partial wave expansion of the amplitude, for each isospin channel, reads
\bea
A_{KK|KK}^I = \frac{32\p}{\rho} \, \sum_{J=0}^\infty \, (2J+1) \, P_J(\cos\theta) \, f_{KK|KK}^{(J,I)}(s) \;,
\label{ps.1}
\eea 
where $ f_{KK|KK}^{(J,I)}$ is the non-relativistic scattering amplitude and 
$ \rho = \sqrt{1- 4\, M_K^2/s} $ .
Our amplitudes are written as 
\bea
A_{KK|KK}^I = A_{KK|KK}^{(0,I)} + (t-u) \, A_{KK|KK}^{(1, I)} + \cdots 
\label{ps.3}
\eea 
In the CM, one has $ (t-u) = (s-4\,M_K^2) \, \cos\theta $
and write
\bea
A_{KK|KK}^I &\!=\!& A_{KK|KK}^{(0,I)} + [(s-4\,M_K^2) \, \cos\theta] \, A_{KK|KK}^{(1, I)} + \cdots 
\nn\\[4mm]
&\!=\!&  \frac{32\p}{\rho} \, \lb f_{KK|KK}^{(0,I)}(s) + 3\, \cos\theta \, f_{KK|KK}^{(1,I)}(s)  + \cdots \rb 
\label{ps.5}
\eea
with
\bea
&& f_{KK|KK}^{(0,I)} = \frac{\rho}{32\,\p}  \; A_{KK|KK}^{(0,I)} \;,
 \label{ps.6}\\[4mm]
&& f_{KK|KK}^{(1,I)} = \frac{\rho^3}{96\,\p}\ \; s\,  A_{KK|KK}^{(1,I)} \;.
\label{ps.7}
\eea
In non-relativistic QM,  the amplitude $f$ is usually expressed \cite{Hyams} in terms of phase shifts 
$\d$ and inelasticity parameters $\eta$ as
\bea 
f_{KK|KK}^{(J,I)} = \frac{1}{2i} \lb \, \eta_{KK|KK}^{(J,I)}  \, e^{2\, i \, \d_{KK|KK}^{(J,I)} } -1 \rb \;.
\label{ps.8}
\eea

In order to obtain $ \lb \d_{KK|KK}^{(J,I)}, \eta_{KK|KK}^{(J,I)}\rb$ from $A_{KK|KK}^{(J,I)}$,
one drops all subscripts and superscripts and write $ f = a + i\, b $,  
with $ a =  \mathrm{Re}\lb f \rb , \, b =  \mathrm{Im}\lb f \rb $. 
Using eq. (\ref{ps.8}), one has 
\bea 
1 + 2\,i\, f &\!=\!& [1 -2\, b] + 2\, i\, a 
= \eta \, \lb \cos 2\d + i\, \sin 2\d \rb 
\label{ps.12} 
\eea 
and thus
\bea
&& \eta = \sqrt{ [1\sm 2\, b]^2 + 4\, a^2}
\label{ps.13}\\[4mm]
&& \tan \d = \frac{2\,a}{1+\eta-2\,b} 
\label{ps.16}
\eea
As $(1+\eta-2\,b) \geq 0$, the sign of $\d$ is determined by the factor $a$.

\section{model structure}
\label{case}

The Multi-Meson-Model we consider in this work assembles a number of aspects that
appear scattered in many calculations, but are normally absent in heavy meson decay analyses.
The main unusual dynamical effects included into our model concern:
i) the presence of a LO contact interaction in the two-body kernel, as 
indicated in Fig.\ref{modA};
ii) the introduction of two resonances in the $(J=0,I=0)$ channel, preserving unitarity;
iii) consideration of coupled channels.
With the purpose of disclosing the role played by these features in the results, 
in this appendix we consider the scattering amplitude $A^{(0,0)}$ and show its behavior in a number of 
different scenarios.
We begin by the simplest one, in which just the $f_0(980)$ is kept,
and add the other contributions gradually, as described in table \ref{scenarios}.
It indicates when a particular contribution, that was previously absent,
has been turned ON.

\begin{table}[h]
\begin{tabular} {|c|c|c|c|c|c|}
\hline
scenario & A 	& B & C & D & MMM \\[2mm] \hline \hline

octet resonance $f_0(980)]$ & \hspace*{3mm} ON \hspace*{3mm} & \hspace*{3mm} ON \hspace*{3mm}  
& \hspace*{3mm} ON \hspace*{3mm}  & \hspace*{3mm} ON \hspace*{3mm}  & \hspace*{3mm} ON \hspace*{3mm} 
 \\ \hline
contact interaction  & x & ON & ON & ON & ON   \\ \hline
singlet resonance $f_0(1370)$ & x 	& x & ON & ON & ON  \\ \hline
 $\pi \pi\, $ coupled channel & x	& x & x & ON & ON  \\ \hline
 $\eta \eta\, $ coupled channel  & x	& x & x & x & ON  \\ \hline
\end{tabular}
\caption{Systematic investigation of the relative importance of $A^{(0,0)}$  components.}
\label{scenarios}
\end{table}

\begin{figure}[h] 
\includegraphics[width=0.48\columnwidth,angle=0]{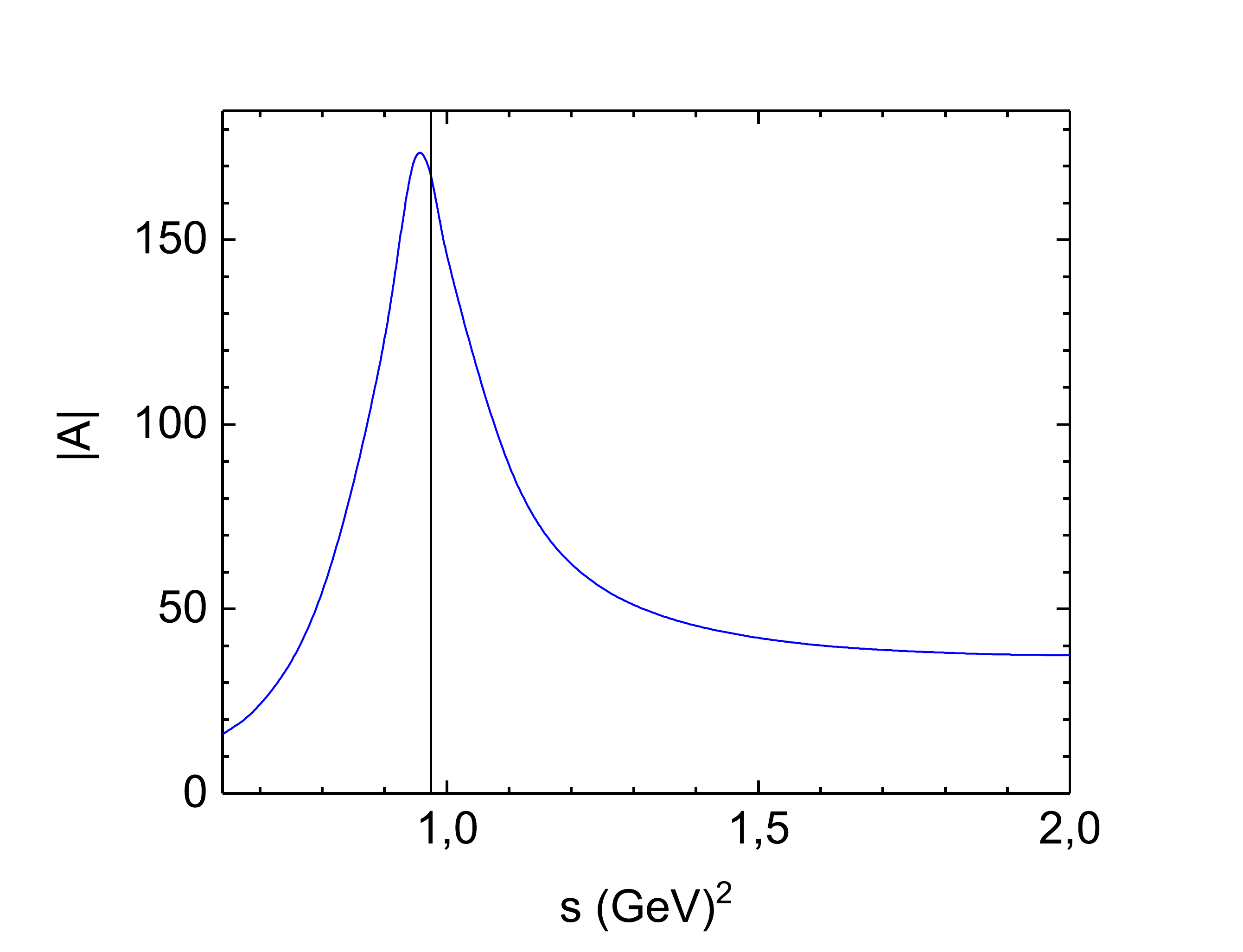}
\includegraphics[width=0.48\columnwidth,angle=0]{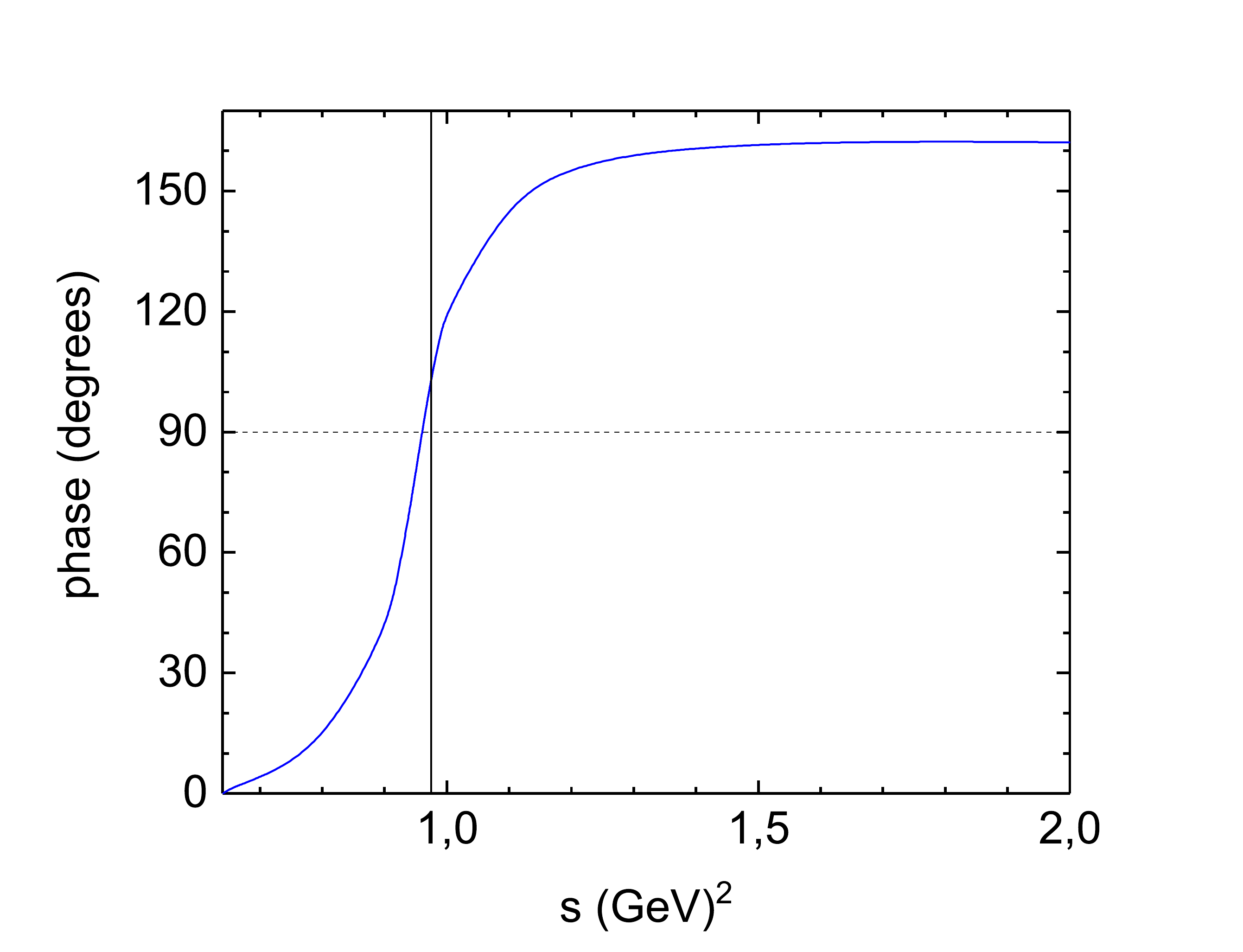}
\caption{Results for $|A^{(0,0)}|$ - the kaon mass is artificially lowered to $M_K=0.4$ GeV
and dynamics is implemented just by a single $f_0(980)$;
the black vertical line indicates the actual $K\Kb  $ threshold.
left: modulus, right: phase.}
\label{fakeK}
\end{figure}

We begin by considering the artificial situation in which the kaon mass is
lowered to $M_K=0.4$ GeV, so as to allow the $f_0(980)$ to be above threshold.
The amplitude is shown in Fig.\ref{fakeK} and results are rather conventional. 
The vertical black line indicates the position of the empirical $K\Kb$ threshold and therefore, 
in actual scattering, one sees only the post-peak part of the resonance,
represented by the blue curves, for scenario A, in Fig. \ref{ScattA}.
Phases in that figure follow general theorems in quantum scattering theory. 
In the absence of inelasticities, the phase of a generic scattering amplitude $A $ 
coincides with the usual phase shift $\delta $ and,
at low energies the latter $\rar 0$ as $ q^{(2L+1)}$, where $ L$ is the 
angular momentum and $q$ is the CM linear momentum.

\begin{figure}[h] 
\includegraphics[width=0.48\columnwidth,angle=0]{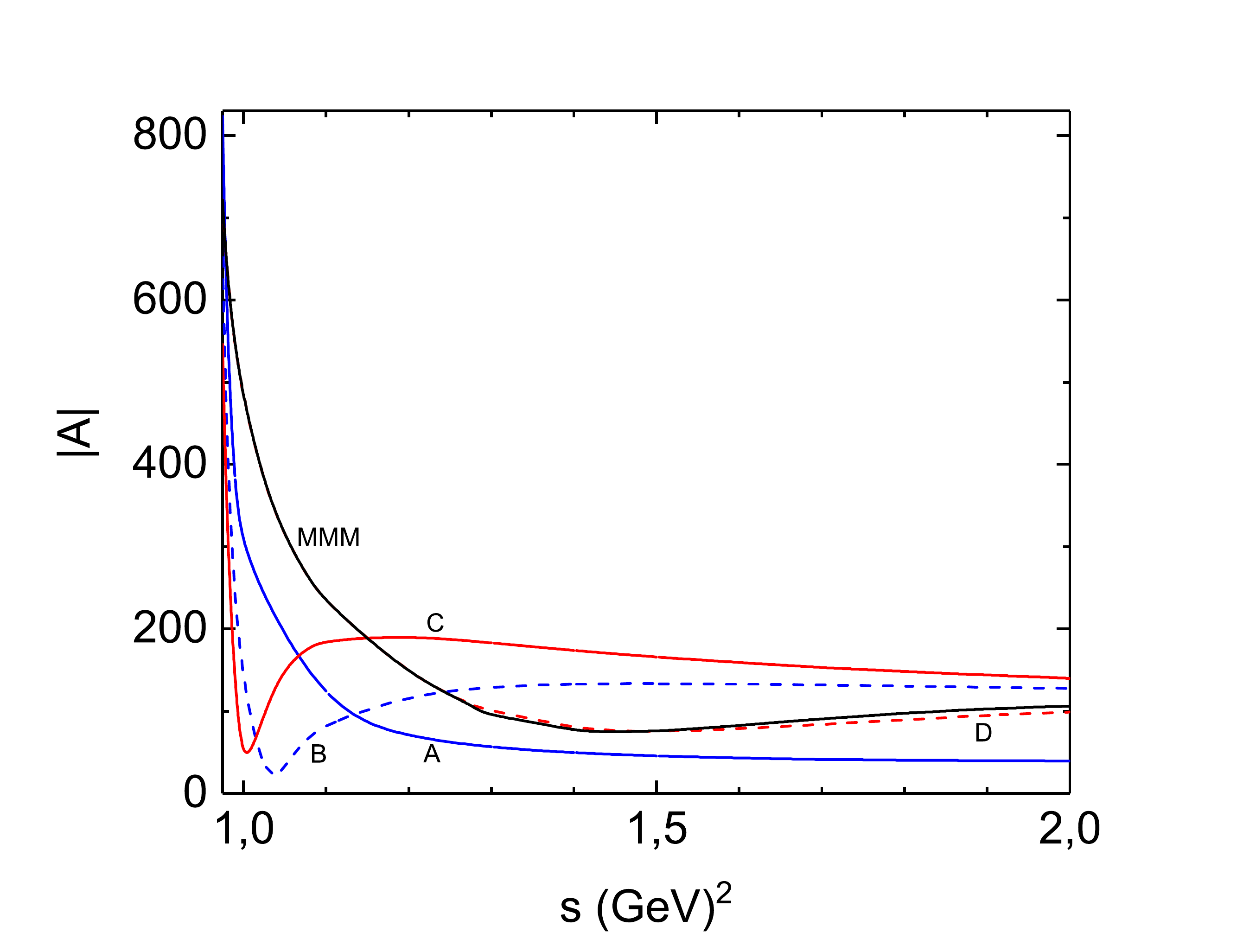}
\includegraphics[width=.48\columnwidth,angle=0]{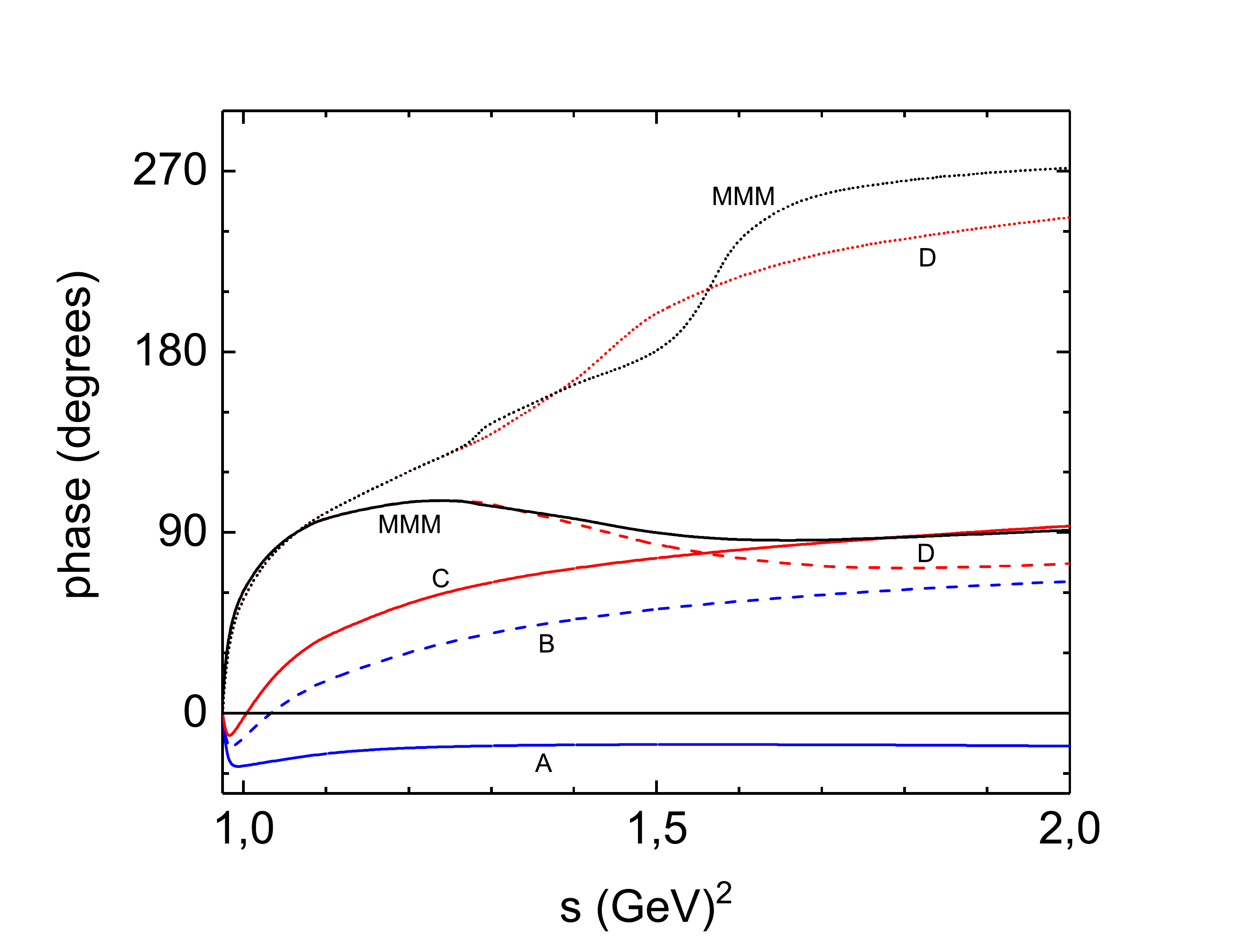}
\caption{Results for $|A^{(0,0)}|$ - piecemeal construction of the amplitude, 
following steps given in table \ref{scenarios};
the continuous blue line (A) corresponds to the tail of the $f_0(980)$;
the dashed blue curve (B) includes the contact chiral term;
the red continuous curve (C) represents the unitarized $f_0(980)$ and $f_0(1370)$ joint contributions;
the dashed red curve accounts for the coupling to $\p\p$ intermediate states;
the continuous black curve (MMM) includes coupling to $\eta\eta$ intermediate states;
top: modulus; bottom: phases; the latter also includes conventional phase shifts $\d^{(0,0)}$,
indicated by the dotted curves. }
\label{ScattA}
\end{figure}

Inspecting these figures, one learns that the inclusion of the chiral contact term (A $\rar$ B)
and the second resonance (B $\rar$ C) produces a strong impact on results.
The influence of the coupling to the $\p\p$ intermediate channel (C $\rar$ D) is  also rather large,
especially at low energies, whereas $\eta\eta$ coupling (D $\rar $ MMM) is much less important.
In Fig.\ref{inel} we show the inelasticity parameter $\eta$.
One must have $\eta = 1$ for elastic amplitudes, and we would like to draw attention to the case
of scenario C, that includes two resonances and no coupled channels. 
In this case, the result for $\eta$ stresses that our method for dealing with multiple 
resonances is indeed consistent with unitarity.
When the coupling to other channels is allowed, $\eta \leq 1$ and the dominance of 
$\p\p$ intermediate states becomes clear.

\begin{figure}[h] 
\includegraphics[width=.48\columnwidth,angle=0]{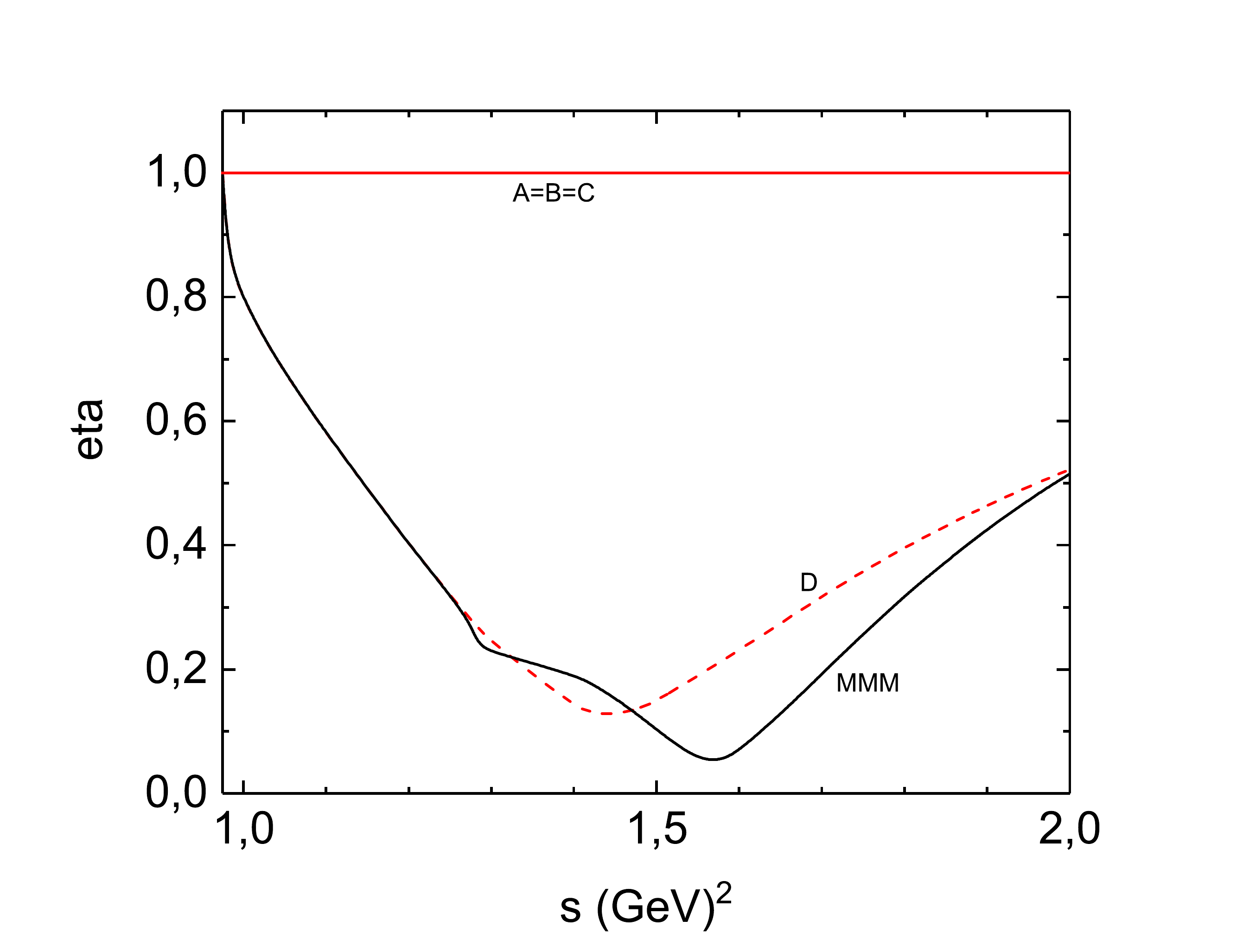}
\caption{Results for $|A^{(0,0)}|$  inelasticities; conventions are the same as in Fig.\ref{ScattA}.}
\label{inel}
\end{figure}



\begin{thebibliography}{99}

\bibitem{sigma} E.M. Aitala {\it et al.}  
(E791), Phys. Rev. Lett.  {\bf 86}  770 (2001);  
E.M. Aitala et al. (E791), Phys. Rev. Lett. {\bf 86} 765 (2001).

\bibitem{E791kappa} E.M. Aitala {\it et al.}  
(E791), Phys.\ Rev.\ Lett. {\bf 89}, 121801 (2002).


\bibitem{LHCb_BW_exemplo} 
R. Aaij {\it et al.} [LHCb colaboration], arXiv:1712.09320, submitted to  Phys. Rev. Lett. 2017 ;
JHEP {\bf 03} 140 (2018); arXiv:1712.08609, submitted to Eur. Phys. J. C. 

\bibitem{BoitoRESUM} D. Boito, J. -P. Dedonder, B. El-Bennich, R. Escribano,
 R.  Kaminski, L. Lesniak and B. Loiseau, 
Phys. Rev. {\bf D96} 11, 113003(2017).

\bibitem{BR} P.C. Magalh\~{a}es, M.R. Robilotta,
K.S.F.F. Guimar\~{a}es, T. Frederico, W.S. de Paula,
I. Bediaga, A.C. dos Reis, 
and C.M. Maekawa, Zarnauskas,G.R.S,
Phys. Rev. D{\bf 84}, 094001 (2011). 

\bibitem{PatWV} P.C. Magalh\~{a}es and M.R. Robilotta, Phys.Rev. D {\bf 92} (2015) 094005.

%
\bibitem{tobias} K.S.F.F. Guimar\~{a}es, W. de Paula, I. Bediaga, A. Delfino, T. Frederico, A. C. dos Reis and L.
Tomio,
Nucl. Phys. B (Proc. Suppl.) {\bf 199} (2010) 341.


\bibitem{kubis} Franz Niecknig and Bastian Kubis, 
 JHEP {\bf 10}, 142 (2015);
  Phys.\ Lett.\ B {\bf 780} (2018) 471.

\bibitem{satoshi} S. X. Nakamura, Phys. Rev. {\bf D 93}, 014005 (2016).


\bibitem{Hyams} B. Hyams et al. Nucl. Phys. B{\bf 64}, 134 (1973).
\bibitem{cohen} Cohen, D. et al., {\sl Phys. Rev.}
{\bf D7}, 661  (1973).

\bibitem{mousallam_KKff} M. Albaladejo, B. Moussallam, Eur. Phys. J. C75(10), 488 (2015), 1507.04526.

\bibitem{mousallam_eta3pi} M. Albaladejo and B. Moussallam, B. 
 Eur. Phys. J. {\bf C77} no8, 508 (2017).

\bibitem{OllerMeissner} Ulf-G. Mei$\beta$ner, J.A. Oller, Nucl. Phys. {\bf A 679} 671 (2001), [arXiv:hep-ph/0005253]. 

\bibitem{TM}
  R.~T.~Aoude, P.~C.~Magalh\~{a}es, A.~C.~dos Reis and M.~R.~Robilotta,
  PoS CHARM {\bf 2016} (2016) 086
  [arXiv:1604.02904].



\bibitem{FOCUS}
J.M. Link {\it et al.}  [FOCUS Collaboration], Phys.\ Lett.\  B {\bf 681},  (2009) 14;

\bibitem{compare}  
D. R. Boito, R. Escribano, Phys. Rev. D {\bf 80}, 054007 (2009);
M. Diakonou and F. Diakonos, Phys. Lett. B {\bf 216}, 436 (1989).

\bibitem{LASS} D. Aston et al., Nucl.Phys. B {\bf 296}, 493 (1988).

\bibitem{Pelaez}  R.~Garcia-Martin, R.~Kaminski, J.~R.~Pelaez, J.~Ruiz de Elvira and F.~J.~Yndurain,
  Phys.\ Rev.\ D {\bf 83}, 074004 (2011)
  [arXiv:1102.2183 [hep-ph]].

\bibitem{OOunit} J.A. Oller and E. Oset, 
Phys. Rev. D {\bf 60}, 074023 (1999);
Nucl. Phys. A {\bf 620}, 465 (1997); A {\bf 652}, 407(E) (1999);
J.R. Pelaez, J.A. Oller  and  E. Oset,
Nucl.Phys. A675,  92C (2000);
K.P. Khemchandani, 
A.  Martinez Torres, H. Nagahiro and A. Hosaka, Phys.Rev. {\bf D88},
114016 (2013).

\bibitem{unit} I.J.R. Aitchison,  e-Print: arXiv:1507.02697;
 J.H.A. Nogueira,  e-Print: arXiv:1605.03889.

\bibitem{EU} G. Ecker and R. Unterdorfer, Eur.Phys.JC{\bf 24}, 535 (2002),
Nucl.Phys.Proc.Suppl. {\bf 121}, 175 (2003).
\bibitem{WChPT} S. Weinberg, Physica A {\bf 96}, 327 (1979).

\bibitem{GL84} J. Gasser and H. Leutwyler, Ann. Phys. {\bf 158}, 142 (1984).

\bibitem{GL85} J. Gasser and H. Leutwyler, Nucl. Phys. B{\bf 250}, 465 (1985).

\bibitem{EGPR} G. Ecker, J. Gasser, A. Pich and E. De Rafael,
Nucl. Phys. B {\bf 321}, 311 (1989).

\bibitem{HM} G. Burdman and J.F. Donoghue, 
Phys. Lett. B {\bf 280}, 287 (1992); M.B. Wise, Phys.Rev. D{\bf 45}, R2188 (1992).

\bibitem{PDG} C. Patrignani et al. (Particle Data Group), Chin. Phys. C, 40, 100001 (2016) 
and 2017 update. 

\bibitem{LHCbkkk} R. Aaij {\it et al.} [LHCb colaboration] LHCb-CONF-2016-008, 
CERN-LHCb-CONF-2016-008. 




%


%
%
 
%
%
%
%
%
%
%
%
%
%
%
%
%
%
%
%
%



\end{thebibliography}
\end{document}